\documentclass[usegraphicx,usenatbib,apj,iop]{emulateapj}

\usepackage{epsfig}
\usepackage{amsmath}
\usepackage{amssymb}
\usepackage{makeidx}
\usepackage{ifthen}
\usepackage{verbatim}
\usepackage{listings}

\usepackage{hyperref}
\hypersetup{
pdftitle={RR Lyrae in XSTPS: The halo density profile in the North Galactic Cap},
pdfauthor={Lorenzo Faccioli},
colorlinks=true,
linkcolor=blue,
citecolor=blue,
urlcolor=blue
}

\begin{document}

\shorttitle{RR Lyrae in XSTPS: The halo density profile in the North Galactic Cap}
\shortauthors{Faccioli et al.}

\def \xdss {XSTPS} 
\def \kpc {\mathrm{kpc}}
\def \deg {\mathrm{deg}}
\def \rgc {R_\mathrm{GC}}
\def \zgc {Z_\mathrm{GC}}

\title{RR Lyrae in XSTPS: The halo density profile in the North Galactic Cap}

\author{
L.~Faccioli\altaffilmark{1,2},
M.~C.~Smith\altaffilmark{3,1,2},
H.-B.~Yuan\altaffilmark{1,4},
H.-H.~Zhang\altaffilmark{5},
X.-W.~Liu\altaffilmark{1,5},
H.-B.~Zhao\altaffilmark{6}, 
and
J.-S.~Yao\altaffilmark{6}
}

\altaffiltext{1}{Kavli Institute for Astronomy and Astrophysics,
Peking University, Beijing 100871, P. R. China}
\altaffiltext{2}{National Astronomical Observatories, 
Chinese Academy of Sciences, Beijing 100012, P. R. China}
\altaffiltext{3}{Key Laboratory for Research in Galaxies and
Cosmology, Shanghai Astronomical Observatory, Chinese Academy of
Sciences, 80 Nandan Road, Shanghai 200030, China; msmith@shao.ac.cn}
\altaffiltext{4}{LAMOST Fellow}
\altaffiltext{5}{Department of Astronomy,
Peking University, Beijing 100871, P. R. China}
\altaffiltext{6}{Purple Mountain Observatory, Chinese Academy of Sciences,
Nanjing 210008, P. R. China}

\begin{abstract}
We present a catalog of RR Lyrae stars (RRLs) observed by the
Xuyi Schmidt Telescope Photometric Survey (\xdss).
The area we consider is located in the North Galactic Cap, covering
$\approx376.75~\deg^2$ at $\alpha\approx150~\deg$ and
$\delta\approx27~\deg$ down to a magnitude limit of $i\approx19$.
Using the variability information afforded by the multi-epoch nature
of our \xdss~data, combined with colors from the Sloan Digital Sky
Survey, we are able to identify candidate RRLs.
We find 318 candidates, derive distances to them and estimate the
detection efficiency. The majority of our candidates have more than 12
observations and for these we are able to calculate periods. These
also allows us to estimate our contamination level, which we predict
is between 30\% to 40\%.
Finally we use the sample to probe
the halo density profile in the $9-49~\kpc$ range and find that it can
be well fitted by a double power law. We find good agreement between
this model and the models derived for the South Galactic Cap
using the \citet{watkins09} and \citet{sesar10} RRL data-sets, after
accounting for possible contamination in our data-set from Sagittarius
stream members. We consider non-spherical double power law models of
the halo density profile and again find agreement with literature
data-sets, although we have limited power to constrain the flattening
due to our small survey area. Much tighter constraints will be placed
by current and future wide-area surveys, most notably ESA's
astrometric Gaia mission. Our analysis demonstrates that surveys with
a limited number of epochs can effectively be mined for
RRLs. Our complete sample is provided as accompanying online
material.
\end{abstract}
\keywords{stars: variables: RR Lyr\ae~ -- Galaxy: halo, stellar content, structure -- techniques: photometric -- surveys}

\section{Introduction}
\label{sec:intro}
Modern theories of galaxy formation and evolution predict that galaxies are formed by
gradual accretion of baryonic matter falling into the gravitational potential of
surrounding cold dark matter haloes (see \citealt{freeman02} for a review).
In this view large galaxies such as the Milky Way grow by accreting material from
smaller nearby galaxies, which are tidally destroyed by the gravitational pull of the 
larger ones.
Over cosmic time these small galaxies are therefore fragmented and reduced to clumps
and streams of stars in the halo of the larger disrupting galaxy;
these streams and clumps can in principle be detected by large scale
surveys of the stellar halo of the large galaxy, allowing the theory to be tested.
The test can be more easily carried out in the Milky Way halo.
In the last decade or so
considerable progress has been made thanks to the unprecedented capabilities of modern
astronomical instrumentation that are now able to survey large portions
of the Milky Way halo providing accurate photometric and spectroscopic data for millions
of its stars.
\par
Foremost among the large scale surveys of the halo has been the Sloan Digital Sky Survey
\citep[SDSS:][]{york00}\footnote{\url{http://www.sdss.org}} whose data-set provides,
among many other things, accurate photometric data in five bands for millions and
spectra for hundreds of thousand of stars in the Milky Way.
This data-set has been successfully used to study the Milky Way in
great detail and on
a large scale, both in the disk \citep[e.g.][]{ivezic08,juric08,bond10}
and in the halo.
With regard to the halo, which is the focus of this paper,
notable examples of use of SDSS data include 
\citet{newberg02}, who identify new structures in the halo Way from
analyzing five million stars, \citet{yanny03}, who
find evidence for a ring of stars in the plane of the Milky Way at
$\approx~18~\mathrm{kpc}$, and \citet{belokurov06}, who use SDSS
photometry to study Milky Way halo substructure in the area
around the north Galactic cap
\par
While these studies probe the structure of the Milky Way via its whole stellar population
and use samples comprised of millions of stars, with all the problems the analysis of
such large data-sets entails, other approaches focus on identifying suitable
tracer populations and use those tracer populations to probe the structure.
When tracers are used the key problem is of course to identify
the tracer itself and show that it indeed traces the underlying
stellar population; the advantage is that one typically deals with samples comprised of
hundreds or thousands of stars as opposed to millions.
\par
Once again SDSS has been instrumental in enabling this kind of analyses: for
example \citet{xue08} use Blue Horizontal Branch (BHB) stars from SDSS to derive the
rotation curve of the Milky Way up to $\approx 60~\kpc$ from the center and \citet{smith09a}
build a sample of $\approx 1700$ halo stars in the solar neighborhood and use it to constrain
the halo velocity dispersion and to look for substructure in the halo, as predicted by
theories of galaxy formation.
\par
Although SDSS has proved very useful for studying the stellar halo, one drawback is
that for the most part it lacks multi-epoch data (the exception to this being the
$\approx 300~\deg^2$ square degree Stripe 82 region; see below).
This means that if one is interested in probing the structure of the Milky Way using a
tracer population of variable objects, the usefulness of SDSS is severely reduced
since with single epoch data alone it is in general not feasible to ascertain the
variable nature of an object
\par
This is an important concern if one wants to use RR Lyrae stars (RRLs) as a halo
tracer population to study the Milky Way stellar halo.
RRLs are Horizontal Branch stars like the BHB stars mentioned above; 
they are old and metal poor like the general halo stellar population and therefore
constitute an excellent tracer.
Their most notable characteristic is that they are variable objects
with a period in the $0.2-1~\mathrm{d}$ range 
but in $V$ they have the same absolute brightness modulo the
effect of metallicity and this fact can be used to derive a distance
to them; 
unlike Cepheids, RRLs are lacking in well calibrated Period-Luminosity
relations, although progress is being made on that front
\citep{caceres08,madore13,klein14}.
For a review of the use of RRLs stars as distance indicators see
\citealt{bono03a}.

They are also abundant and relatively easy to find if accurate
multi-epoch data are available, as shown by the microlensing
experiments in the 1990s. The three main microlensing experiments
\citep[MACHO:][]{alcock93}, \citep[OGLE:][]{udalski92}, \citep[EROS:][]{aubourg93}
proved to be invaluable
for the study of variable objects in the Magellanic Clouds and the Milky Way bulge
thanks to the hundreds of epochs available in their data-sets for each object; large
catalogs of RRLs feature prominently among the data products of these surveys
\citep{alcock98, soszyinsky09, soszyinsky10, soszyinsky11}.
The key to finding RRLs in large numbers in these data-set is the availability of
hundreds of epochs, which enables to unambiguously establish the variable nature of 
an object and to find the RRL period with great precision;
this however is not possible with single (or few) epoch data-sets such as SDSS.
\par
In a pioneering paper \citet{ivezic05} show how RRLs can be reliably identified in the SDSS
data-set by a combination of color cuts, even with few (or just one) epochs;
in particular the cuts they propose are:
\begin{eqnarray}
\label{eq:cuts}
\nonumber
14 < r &<& 20, \\
\nonumber
0.98 < u-g &<& 1.3, \\
\nonumber
D_{ug}  \equiv (u-g)+0.67(g-r) &-& 1.07, \\
\nonumber
D_{gr}  \equiv 0.45(u-g)-(g-r) &-& 0.12, \\
\nonumber
D_{ug}^{\mathrm{Min}} < D_{ug} &<& 0.35, \\
\nonumber
D_{gr}^{\mathrm{Min}} < D_{gr} &<& 0.55, \\
\nonumber
-0.15 < r-i &<& 0.22, \\
-0.21 < i-z &<& 0.25
\end{eqnarray}
where $D_{ug}^{\mathrm{Min}}$ and $D_{gr}^{\mathrm{Min}}$ can be
chosen to give a desired selection completeness and efficiency.
\citet{ivezic05} however show that, for a complete RRL sample, their
selection efficiency is only $6\%$ (this efficiency may be increased
at the price of reduced completeness with different cuts), that is,
in a color-selected sample that includes all RRLs, these will make
up just $6\%$ of the selected objects.
It is therefore necessary that studies of the Milky
Way halo based on RRLs should include criteria other than
the cuts in Equations \ref{eq:cuts} to cleanly select a contamination
free sample; such criteria usually exploit the availability of
multi-epoch and multi-band photometry (though much fewer epochs than
the microlensing searches are needed).
\par
While most SDSS fields have been in general observed once or few
times, this is not the case for a particular region in the South
Galactic Cap (SGC) known as Stripe
82\footnote{\url{http://www.sdss.org/legacy/stripe82.html}}
which has accurate five band and multi-epoch photometry.
SDSS Stripe 82 has therefore been the object of much recent work
aimed at efficiently detecting variable objects, including RRLs, using combinations of color
cuts and variability information provided by sparsely sampled
light-curves (typically objects in Stripe 82 have light-curves
with at most a few tens of points, as opposed to the several
hundreds or even thousands of points in microlensing searches).
\citet{sesar07} investigate the problem of detecting variable
objects with just as few multi band photometric measurements as
four and propose a series of criteria they apply to variable
objects in Stripe 82.
\par
Using these criteria and the cuts in Equations \ref{eq:cuts} with
$D_{ug}^{\mathrm{Min}}=-0.05~\mathrm{mag}$ and
$D_{gr}^{\mathrm{Min}}=0.06~\mathrm{mag}$ \citet{sesar07} also select 683
RRL candidates from Stripe 82 and use it to investigate the structure of the Milky Way halo down to 
$\approx 100~\kpc$; however this sample still suffers from
contamination due to the generally low number of available
observations.
\par
\citet[][hereafter W09]{watkins09} use the same set of color cuts, with the
same $D_{ug}^{\mathrm{Min}}$ and $D_{gr}^{\mathrm{Min}}$ and slightly
different criteria for detecting variability, and look for RRLs
in Stripe 82 using the light motion catalog presented by \citet{bramich08}.
W09 find 407 RRLs (316 RRab and 91 RRc) and detect over-densities in the Milky Way halo,
including a newly discovered one at $\approx 80~\kpc$, $l\approx 80~\deg$,
$b\approx -55~\deg $ which they term the Pisces Overdensity.
\par
Using additional data in Stripe 82 from the SDSS Supernova Survey
\citep{frieman08,sako08}\footnote{\url{http://www.sdss.org/supernova/aboutsprnova.html}}
\citet[][hereafter S10]{sesar10} update the analysis performed in \citep{sesar07} and present a new catalog composed of
483 objects; S10 note that, while the main conclusions of
\citet{sesar07} are still valid, only $\approx 70\%$ of
\citet{sesar07} RRL candidates turn out to be true RRL in
S10, with the majority of the other candidates turning
out to be $\delta$ Scuti stars.
By taking advantage of the high temporal cadence afforded by the SDSS Supernova Survey S10
are also able to construct detailed light-curve templates in the five SDSS bands; this improved sampling
also allows S10 to relax their color cuts from $u-g>0.98~\mathrm{mag}$ to
$u-g>0.7~\mathrm{mag}$ to better find low
metallicity RRLs.    
\par
Several RRL samples not using SDSS data have also been published in
recent years.
The QUEST RRL survey \citep{vivas04} found 498 RRLs in a $380~\deg^2$
area from $4^{\mathrm h}.1$ to $6^{\mathrm h}.1$ and
from $8^{\mathrm h}$ to $17^{\mathrm h}$ at $\delta\approx -1~\deg$, in the
$13.5~\mathrm{mag}<V<19.5~\mathrm{mag}$ range, which
allows them to probe the structure of the halo from $\approx 4~\kpc$ to
$\approx 60~\kpc$.
Using this data-set \citet{vivas06} investigate over-densities in
the halo: they find that away from major over-densities the halo
can be well fitted by a smooth, non-spherical (oblate) profile
and use this profile to estimate a ``background'' and to look for
over-densities overlaid on top of it.
\par
\citet{kinemuchi06} present a sample of 1197 candidate RRLs in the
solar neighborhood from the Northern Sky Variability Survey (NSVS).
\citet{keller08} present 2016 candidate RRLs from the
Southern Edgeworth-Kuiper Belt Object \citep{moody03} (SEKBO)
survey up to
$\approx~50~\mathrm{kpc}$ and map halo over-densities,
revealing a series of structures coincident with the leading and
trailing arms of debris from the Sagittarius dwarf galaxy.
\citet{akhter12} use the \citet{keller08} sample to study the
steepening of the RRL distribution power-law slope beyond a
Galacto-centric radius of $\approx~45~\mathrm{kpc}$ and
find evidence for a change from $2.78\pm 0.02$ in the inner
halo to $-5.0\pm 0.2$ in the outer.
\citet{miceli08} present a sample of 838 RRLab from the Lowell
Observatory Near Earth Objects Survey Phase I
\citep{bowell95} (LONEOS-I) up to
a galacto-centric distance of $30~\mathrm{kpc}$ and find evidence
for dual-halo models of halo formation.

While this work was in preparation, two new catalogs of RRL have
been assembled.
Firstly the LINEAR survey \citep{sesar11a} has been mined,
resulting in a sample of around 5000 RRL
\citep{sesar13,palaversa13}. This catalog was then used to
investigate the properties and distributions of the various types of
RRL, measuring the halo profile and searching for substructure in the
nearby (5-30 kpc) stellar halo.
Secondly, \citet{drake13a} has presented a sample of 12227 type-ab
RRLs covering $\approx 20,000~\deg^2$ to heliocentric distances up to $60~\kpc$
assembled from public light-curves in the Catalina Survey
Data Release 1\footnote{\url{http://crts.caltech.edu/}}, providing periods and
distances, and revealing parts of the Sagittarius Stream at heliocentric
distances $20$ to $60~\kpc$.  
\citet{drake13b} use 1207 RRLs found in photometry from the Catalina Survey's
Mount Lemmon Telescope to detect RRLs up to $100~\kpc$ and find evidence for
a tidal stream beyond $100~\kpc$ overlapping the Sagittarius Stream
system. In the near future many other surveys will be able to provide
large catalogs of RRLs, for example the Palomar Transient Factory
(PTF; \citealt{rau2009}), Pan-STARRS \citep{kai2002}
and Gaia \citep{bono03b,eyer12}.
\par
In this paper we present a catalog of 318 candidate RRLs observed by
the Xuyi Schmidt Telescope Photometric Survey of the Galactic
Anti-Center (XSTPS-GAC, \xdss~for short from now on) and use this
catalog to probe the halo density profile in the North Galactic Cap
(NGC) up to $\approx 50~\kpc$.
The survey area is covered by the \citet{drake13a} area so it is likely
that (part of) the RRLs we present are also detected by them.
The paper is organized as follows: Section \ref{sec:xdss}
introduces the \xdss~survey, Section \ref{sec:rrl} describes
our cuts for finding RRL candidates, Section \ref{sec:eff}
describes our efficiency estimation for finding the candidates,
Section \ref{sec:period} describes our period finding procedure
for those candidates with enough observations and also our sample
contamination estimate, Section \ref{sec:dist} describes our
distance estimation.
In Section \ref{sec:solar} we use the \xdss~RRLs to investigate the
halo density profile in the North Galactic Cap in the $10-49~\kpc$
range, and the W09 and S10 RRLs to do the same for the
South Galactic Cap and compare the results; Subsection
\ref{subsec:spherical} deals with spherical halo models
and Subsection \ref{subsec:oblate} deals with non-spherical
halo models.
Finally Section \ref{sec:conc} reports our conclusions.
\par
We point out that in this paper we do not use any period-luminosity
relationship to estimate distances to our RRLs, because the periods
for many of our RRLs are poorly constrained due to having few
observations in just one band. As pointed out above, progress is being
made improving RRL period-luminosity relationships and some are now 
becoming available that can be applied to our catalog
\citep[for example][]{caceres08}. In Section \ref{sec:dist} we
quantify the uncertainty introduced by our distance estimation.
\section{The Xuyi Schmidt Telescope Photometric Survey of the Galactic
Anti-Center (\xdss)}
\label{sec:xdss}
\xdss~is a three band photometric survey of the Milky Way anti-center
carried out with the $1.2~\mathrm{m}$ Schmidt telescope at the Xuyi
station of the Purple Mountain Observatory in China \citep{zhang13}
and completed in March 2011 \citep[see Section 3 of][]{liu13}. The
survey covers an area of $\approx 6,000~\deg^2$ in a region of the sky
that, for the most part, is not covered by other large scale surveys.
It is expected to detect $\approx 100$ million stars in
$i$ and the first data release is due imminently (Yuan et al., in
preparation); we used a preliminary version of the catalog for our study.
The three bands employed in the survey are $g$, $r$, $i$, similar, but not identical, to the
corresponding SDSS bands; the transformations between the \xdss~and the SDSS
systems are given by Equations \ref{eq:transf} and are valid both for
reddened and unreddened magnitudes \citep{yuan13}:
\begin{align}
\label{eq:transf}
g^\mathrm{SDSS}&=& g + 0.144(g-i),  \notag \\
r^\mathrm{SDSS}&=& r + 0.0680(r-i), \notag \\
i^\mathrm{SDSS}&=& i + 0.0224(g-i).
\end{align}
The exposure time is $90~\mathrm{sec}$ and the magnitude limit is
$\approx 19 \mathrm{mag}$ in $i$.
The data used in this work are based on extended and multi-epoch observations
toward the NGC during December-February 2010 and 2011; the extended survey area
covers the range $119.7~\deg\lesssim\mathrm{RA}\lesssim 193.5~\deg$ in Right
Ascension (RA) and $24.6~\deg\lesssim\delta\lesssim 29.7~\deg$ in Declination
($\delta$) for an area $\approx 376.75~\deg^2$.
Most of the observations were carried out in the $i$ band.
\par
\begin{figure}
\begin{center}
\plotone{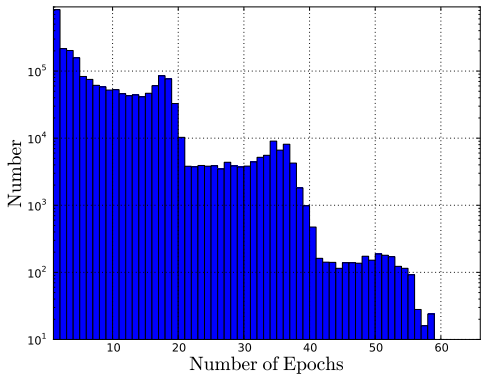}
\end{center}
\caption{Histogram of the number of epochs in the $i$ band.}
\label{fig:epochs}
\end{figure}
\begin{figure}
\begin{center}
\plotone{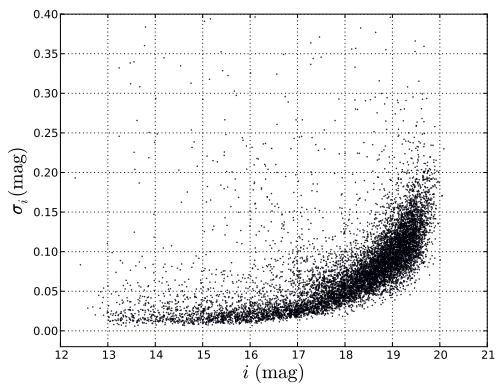}
\end{center}
\caption{RMS $i$ magnitude $\sigma_i$ of 10,000 randomly selected objects
with at least 10 epochs as a function of their mean $i$ magnitude.}
\label{fig:rms_mag}
\end{figure}

Observations in the same band at different epochs were matched using a
matching radius of $3~\mathrm{arcsec}$ \citep[for a discussion of
the XSTPS astrometric accuracy, see][]{zhang14}.
The procedure yielded $\approx 4.2$ million matches in $i$, as well as
$\approx 1.17$ million in $r$ and $\approx 0.5$ million in $g$,
however many of these matches are not true objects but false
sources that must be excluded.
To filter out these false sources we required that, to be considered
as a true object, a source must be detected in at least two epochs; this
leaves us with $\approx 1.6$ million sources that we consider bona fide
objects; while this cut may appear too severe (after all there are
true sources observed in just one epoch), it has no consequences for the
subsequent analysis, in which we will be interested in multi-epoch
objects anyway.
The histogram of the number of epochs for all the 4.2 million sources in the
$i$ band (including the false ones) is shown in Figure \ref{fig:epochs};
the high number of sources having one or two epochs is due to the
aforementioned contamination from false sources.
Figure \ref{fig:rms_mag} shows the RMS $i$ magnitude $\sigma_i$ of
10,000 randomly selected objects with at least 10 epochs as a function
of their mean $i$ magnitude. The figure shows that at $i \approx
19~\mathrm{mag}$, $\sigma_i\approx 0.1~\mathrm{mag}$; since typical
RRL amplitudes can be as low as $0.3~\mathrm{mag}$ \citep{sesar10},
which corresponds to a RRL Amplitude-RMS magnitude ratio of just 3 at
$19~\mathrm{mag}$, we take $19~\mathrm{mag}$ in $i$ as our
completeness limit for the purpose of finding RRLs, consistent with
the survey magnitude limit.

\section{Finding RRLs in \xdss}
\label{sec:rrl}
The method we used to select RRL candidates relies on a
combination of SDSS colors as described by Equations
\ref{eq:cuts} and variability information provided by the
\xdss~data in this stripe.
Therefore the first step is to select stars from SDSS in the region
covered by the \xdss~stripe and cross correlate this sample with the
\xdss~data. We selected objects in the SDSS Data Release 7
(DR7) \footnote{\url{http://www.sdss.org/dr7/}}
from the ``Star'' table in the region
$110~\deg<\mathrm{RA}<200~\deg$ and $24~\deg<\delta<30~\deg$
(J2000), with PSF $i_\mathrm{SDSS}$ magnitudes between $13.5$ and
$22~\mathrm{mag}$.
The query returned 2,445,651 objects that were correlated with the
\xdss~$i^\mathrm{SDSS}$ band catalog in a $3~\mathrm{arcsec}$
radius; this gave $\approx 880,000$ matches for all the
4.2 millions of sources in the \xdss~data-set, including the false
ones; we do not care about those at this stage since they will be
filtered out later.
We applied to these matches the cuts defined by Equations \ref{eq:cuts}
with $D_{ug}^{\mathrm{Min}}=-0.05~\mathrm{mag}$ and
$D_{gr}^{\mathrm{Min}}=0.06~\mathrm{mag}$
\citep{sesar07} and found 12,992 stars with colors satisfying these
these RRL cuts.
\par
The next step in the process is to identify variable sources in
the \xdss~data-set.
Candidate variable sources were selected by first identifying
objects with at least four epochs in the $i$ band (which yielded
$\approx 1.1$ million objects and filtered out the false sources
mentioned above, which usually have just one detection) and then fitting a
constant to these objects and computing the $\chi^2$ fit statistic.
Objects whose probability of having a $\chi^2$ equal to or higher than the one
in the fit was $< 0.001$ were considered potential variables;
this cut yielded 19,556 objects.
One particular concern with this procedure (especially so considering
the low number of observations usually available) is that the
$\chi^2$ may be high because of one single (or very few) discrepant point(s)
due to bad observations.
This was accomplished by refitting all the 19,556 candidate variables
after removing each point in turn: if the low value of $\chi^2$
is due to a single point away from the others, a fit to a
constant should give a good value of $\chi^2$
(defined as having a $0.2$ probability of $\chi^2$ being equal to or
higher than the one in the fit) when this point is removed.
In this case we may establish that the source is either a non RRL
variable or is a constant with a bad point and we can exclude
it from further consideration.
Our cut removed $5,484$ sources from the list of candidate variables
leaving $14,072$ objects that we consider bona fide variables.
There is a chance that stars with a large number of epochs may have
multiple bad observations, but this is dealt with later in Section
\ref{sec:period}.
\par
We finally cross correlate the candidate RRLs found from the color cuts in
Equations \ref{eq:cuts} with the sample of variables found via the criteria
outlined in the above paragraph; this cross correlation finally yielded
318~objects that make up our candidate RRL sample.
This sample includes a few objects with $i$ greater than
our magnitude limit ($19~\mathrm{mag}$); these objects will not be used
in subsequent analyses.
\par
\begin{figure}
\begin{center}
\plotone{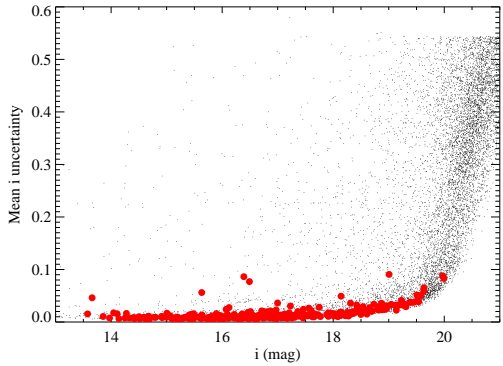}
\end{center}
\caption{Uncertainty on $i$ magnitude for our 318 RRL candidates
(red points) and for 10,000 randomly selected objects in $i$
(small black points).}
\label{fig:err_vs_mag}
\end{figure}

Our procedure for finding RRLs relies on detecting a poor fit
to a constant baseline, so it is dependent on magnitude uncertainties
since an object with a large uncertainty will be less likely to have a
poor fit to a constant.
It is therefore interesting to compare the
uncertainty distribution as a function of magnitude of our
RRL candidates with that of the general population of
objects in the $i$ band from which the RRL sample is
extracted.
Figure \ref{fig:err_vs_mag} shows mean $i$ uncertainty
versus $i$ magnitude for our RRL candidates and for 10,000
randomly chosen objects detected in the $i$ band.
The figure shows that the two distributions are compatible
down to our magnitude limit $i=19$ mag; beyond that the general
$i$ population has a much higher mean magnitude uncertainty
than our RRL candidates, which, in view of our RRL finding
procedure is not surprising.
Even for magnitudes brighter than $19$ however there are several
candidates with large magnitude uncertainty; this has an impact
on detection efficiency as will be shown in Section \ref{sec:eff}.
\par
Complete information for the sample (positions, magnitudes, SDSS matches)
is given in accompanying electronic tables; an example of the data provided
is given, for a few RRL candidates, in Table \ref{tab:data}.
\begin{table}
\caption{Basic data for a few RRLs candidates (see accompanying
online material for full table).}
\begin{tabular}{c c c c c c}
\hline
ID & RA & Dec & Epochs & Mean $i$ & $i$ \\
& & & & (Uncertainty) & Std. Dev.\\
\hline 
10 & 08:36:50.28 & 27:20:24.44 &  9 & 14.58(0.01) & 0.06\\
11 & 08:37:33.25 & 26:51:46.57 &  5 & 17.38(0.02) & 0.19\\
13 & 08:45:15.27 & 27:56:27.89 & 12 & 13.55(0.05) & 0.30\\
15 & 08:47:36.72 & 27:59:58.27 &  7 & 15.99(0.01) & 0.18\\
16 & 08:50:32.88 & 29:30:30.13 &  4 & 14.93(0.02) & 0.15\\
17 & 08:52:19.51 & 27:15:03.04 &  4 & 16.91(0.02) & 0.22\\
18 & 08:54:12.61 & 26:28:39.12 & 10 & 16.77(0.01) & 0.09\\
19 & 08:55:21.58 & 27:53:02.69 &  7 & 19.16(0.04) & 0.20\\
\hline
\end{tabular}
\label{tab:data}
\end{table}

\par

\subsection{Cross correlation with other existing data-sets}
\label{sec:cross_match}
Although it is likely that at least part of our RRLs are found by
\citet{drake13a} we do not expect most of them to have been
identified in the past; to check this we cross-correlated our
RRLs with existing data-sets; a matching radius of
$3~\mathrm{arcsec}$ was used in all searches.
\par
We first cross-correlated our candidate RRLs with the VSX catalog
\footnote{\url{http://www.aavso.org/vsx/index.php}} and found 114
matches, all of which have a period;
the list of the matches, with VSX period, is reported in an accompanying
electronic table.
Of these 114 matches 99 are reported as type-ab RRL (RRLab), 13 as
type-c (RRLc), and only 2 as non RRL.
\par
We then considered other data-sets finding fewer matches.
A cross-correlation of our candidate RRLs with the General
Catalog of Variable Stars
database \footnote{Samus N.N., Durlevich O.V., Kazarovets E V., Kireeva N.N., 
Pastukhova E.N., Zharova A.V., et al.: General Catalog of Variable Stars 
(GCVS database, Version 2012Jan) \url{http://www.sai.msu.su/gcvs/gcvs/index.htm}}
does not yield any match.
Correlating our RRLs with the Strasbourg Astronomical
Data Centre database\footnote{\url{http://cds.u-strasbg.fr/}}
yielded 48 matches of which 27 classified as RRLs;
the remaining either do not have a classification or are
classified as Horizontal Branch stars, as RRLs indeed are.
Correlating the RRLs with available data-sets at
SkyDOT\footnote{\url{http://skydot.lanl.gov/}}, namely the LINEAR
Survey Photometric Database \citep{sesar11a}\footnote{\url{https://astroweb.lanl.gov/lineardb/}}, yielded
just three objects.
No objects were found correlating our RRLs with the
All Sky Automated Survey
(ASAS)\footnote{\url{http://www.astrouw.edu.pl/asas/}}
catalog of 50,122 variable stars, mostly in the Southern
Hemisphere: only 206 ASAS variables are found in the
\xdss~stripe and none of them matched any of our RRLs.
\par
We finally correlated our RRLs with the catalog from
the Digital Access to a Sky Century at Harvard (DASCH)
project \footnote{\url{http://dasch.rc.fas.harvard.edu/}}.
The catalog so far has only partial coverage with the
\xdss~stripe and we found 147 matches.
In most cases a classification for the
matching object was not available;
when it was the object was always classified as an RRL
and we found 24 matching RRLab and 3 RRLc.
\section{Efficiency}
\label{sec:eff}
It is important to quantify the efficiency of our search.
We define the efficiency as the probability
for an RRL with a given mean magnitude and mean uncertainty (taken to be
equal the measured ones)
to be detected by our procedure; this probability depends on a
number of factors.
\par
An important factor is that the number of epochs available for objects
in the XSTPS catalog can vary significantly, as can be seen from
Figure \ref{fig:epochs}. Many objects do not
have enough observations (4 in our case) to allow us to test for variability
and, in general, the higher the number of observations, the more robust our
detection will be.
Furthermore the number of observations is dependent on the position in
the sky, meaning that the probability of an object to have enough
observations to be included in our variability search is dependent on
its position.
\par
Other factors affecting the efficiency are the RRL amplitude and absolute
magnitude (which are also correlated).
\par
To address this concern we used a Monte Carlo (MC) approach to estimate an
efficiency $0\leq\epsilon\leq 1$ for each RRL candidate, with $\epsilon$ depending on
position on the sky.
For each candidate we built 1,000 fake light-curves with a number of epochs drawn from
the local distribution of number of epochs of all objects in a 10 arc-minute box centered at the
candidate position.
If the number of epochs was too low (less than 4) we did not proceed any further; otherwise
a fake light-curve would be built using the templates of S10, following these steps:
\begin{enumerate}
\item
Draw fake times of observation from the actual \xdss~times of
observation distribution.
\item
Draw a fake period from the S10 period distribution and fold the fake times
of observation around this period, getting a phase for each fake time of observation.
\item
Draw a random template from the S10 templates, and an RRL
amplitude from the S10 amplitude distribution and rescale this template by this
amplitude.
The S10 templates are normalized so that their maximum variation is
$1~\mathrm{mag}$; the RRLs in the S10 sample are fitted to a template rescaled
by an amplitude; we drew our fake amplitudes from this amplitude distribution.
Since period and amplitude are correlated, the amplitude was drawn from those
S10 RRLs whose period falls within $\pm 0.01~\mathrm{d}$ of our random period.
\item
Take the RRL mean $i$ magnitude and shift the amplitude-rescaled
template mentioned above so that its mean $i$ magnitude (after transforming
from SDSS to \xdss~magnitudes using the inverse of Equations \ref{eq:transf})
matches the observed $i$ magnitude.
\item
Derive a fake magnitude for each fake time of observation by spline interpolating the rescaled
templates at the phase each time of observation corresponds to.
\item
Use the RRL mean $i$ uncertainty as uncertainty for the fake light-curve.
\end{enumerate}
This procedure gave us a set of fake light-curves to which we applied
the same criteria used to detect variability in our candidates, thus
allowing us to estimate $\epsilon$ as the ratio of the number of MC
realizations passing the cut over the total number of realizations; if
the number of epochs was too low the realization was deemed not to
have passed the cut thus lowering the efficiency.

This technique relies on the assumption that the input templates from
S10 span the entire range of RRL light-curve morphologies and the
accompanying period distribution is unbiased. There are circumstances
where this might not be the case, for example because
magnitude-limited surveys are less likely to detect low-amplitude (and
hence long-period) RRLs.
Although there are such potential flaws, we believe that the high
quality of the S10 catalog should mean that our method provides a good
approximation of the detection efficiency.

\begin{figure}
\begin{center}
\plotone{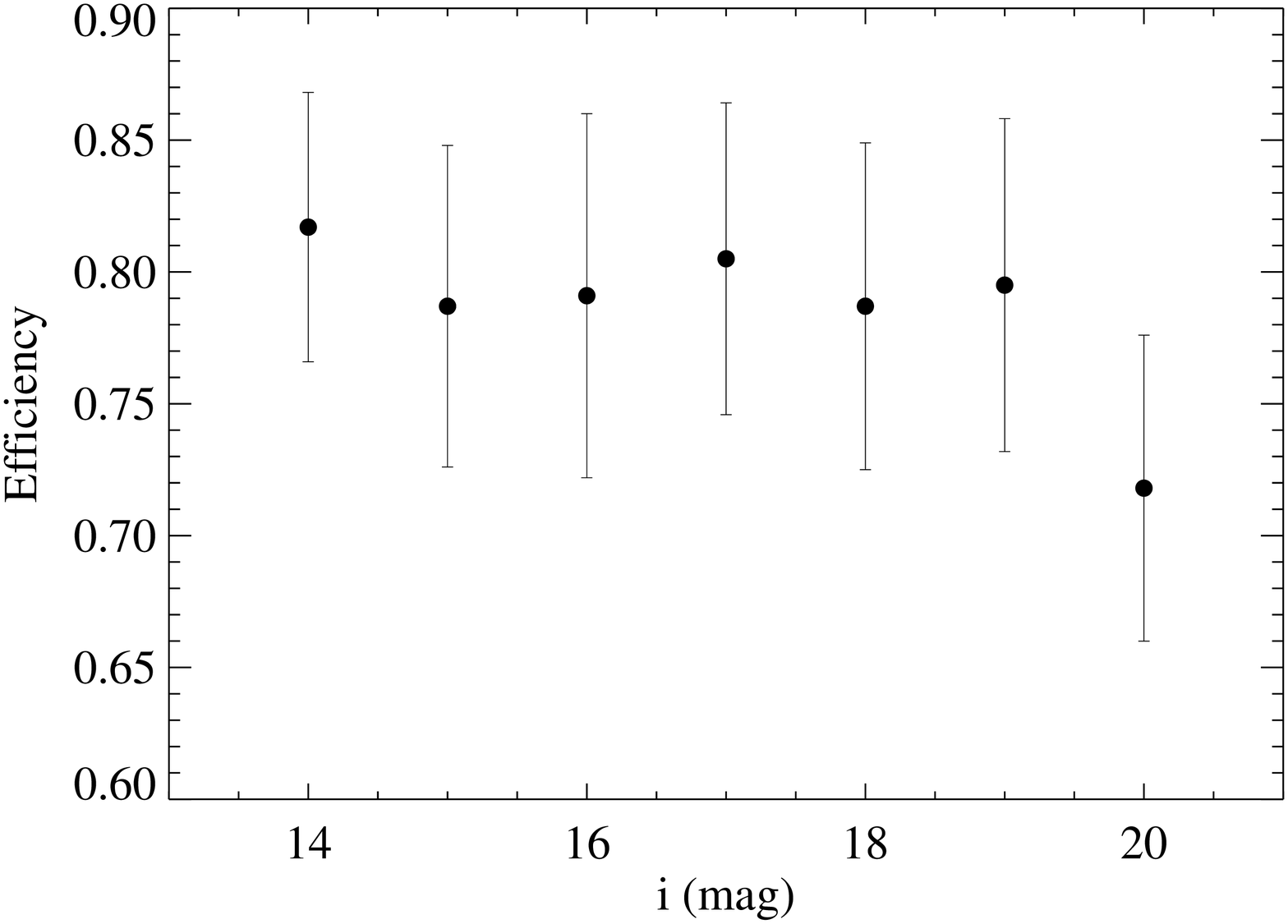}
\end{center}
\caption{Mean efficiency $\epsilon$ (in 1 mag bins) vs. $i$
uncertainty for the 318~RRL candidates. Error bars denote the standard
deviation in each bin.}
\label{fig:eff}
\end{figure}

Our efficiencies are reported in an accompanying electronic table;
figure \ref{fig:eff} shows the mean efficiency in magnitude bins;
efficiency stays at $\approx 80\%$ down to our magnitude limit
(19 mag) and starts declining beyond the limit.
\subsection{Efficiency as a function of magnitude}
The efficiencies show very little dependence on $i$ magnitude down to
our magnitude limit; this behavior requires some explanation.
The aim of our efficiency calculation was to compute the probability
for an RRL with mean magnitude and mean uncertainty equal the measured ones
to be detected by our procedure; four factors conspire
to lower this probability and their dependence on magnitude is minor.
\begin{enumerate}
\item
Not all the fields in the \xdss~strip are observed at least four
times (the minimum number of epochs we required to determine that an
object is variable) so each RRL has $\epsilon<1$ simply because of this.
This factor is obviously independent of magnitude.
\item
A small RRL amplitude makes detecting variability more difficult.
This effect is dependent on magnitude because the amplitude
is correlated with the absolute magnitude, but it is independent of distance.
Replacing the real RRL with a simulated one having a different amplitude
(and therefore a slightly different absolute magnitude) will influence
its detection probability (e.g. making a detection less likely if the
simulated RRL has a small amplitude) in the same way regardless of its
distance (and therefore relative magnitude).
\item
From the point above, there may be a slight dependence of detection
probability on absolute magnitude.
From S10 one sees that the spread in $i$ absolute magnitudes is
$\approx 0.2~\mathrm{mag}$ for both RRLab and RRc, so this is the
maximum spread one can expect between the real and the simulated
RRLs at the same distance; the effect in the efficiency calculation is minor
as will be shown below.
\item
The remaining factor that impacts our efficiency
calculation is the assumed uncertainty in the simulated light-curves,
which we take to equal the measured mean uncertainty.
This uncertainty is in general larger for fainter magnitudes making detecting
variability more difficult.
Our magnitude-uncertainty relation shows the expected trend
(see Figure \ref{fig:err_vs_mag})
but with a large scatter (some bright objects have a large uncertainty and
correspondingly low efficiency) which explains the large error
bars in Figure \ref{fig:eff}; in particular, across a
$\approx 0.2~\mathrm{mag}$ span
(our estimated maximum spread between real and simulated RRLs) the
change in uncertainty is negligible so assuming a simulated uncertainty equal to
the measured one is a good approximation.
\end{enumerate}
Figure \ref{fig:err_vs_mag} also explains why the drop in
efficiency beyond our magnitude limit is minor.
Beyond the magnitude limit our procedure selects RRL candidates
that have, on average, lower uncertainty than the general $i$ band
population and so are probably not a fair tracer of this
population.
We point out again that, since RRL candidates beyond $i=19$ mag
are not used in subsequent analyses, the fact that they may not
be a fair tracer of the $i>19$ mag population is unimportant.

\begin{figure}
\begin{center}
\plotone{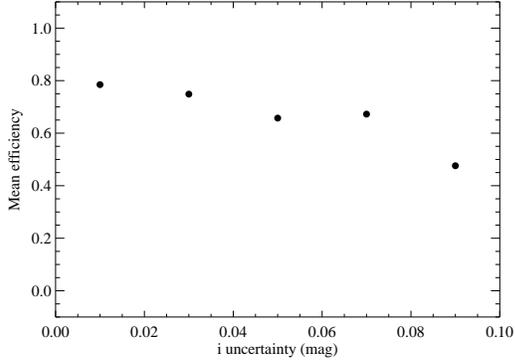}
\end{center}
\caption{Mean efficiency $\epsilon$ (in 0.02 bins) vs. $i$
uncertainty for the 318~RRL candidates.}
\label{fig:eff_vs_err}
\end{figure}

The points above explain the seemingly counterintuitive result of a
detection efficiency almost independent of magnitude.
The main magnitude-dependent factor impacting the efficiency is the
photometric uncertainty, which is to be expected as lower amplitude
RRLs will be harder to detect for lower signal-to-noise
light-curves. This causes the weak drop at fainter magnitudes and is
further illustrated in Figure \ref{fig:eff_vs_err}, which shows mean
$\epsilon$ vs. $i$ uncertainty in $0.02$ uncertainty bins. The trend
of decreasing efficiency with increasing uncertainty is evident.
\section{Period Finding, Light-Curve Folding and Contamination}
\label{sec:period}
Contamination 
(the fraction of non RRL objects that nevertheless pass our color and
variability cuts)
in our sample is an important concern: \citet{sesar07}
show that $30\%$ of their RRL candidates, selected on the basis of
Equations \ref{eq:cuts} and with a median of 10 observations, turn out
not to be RRLs; W09 and S10 on the other hand are able to select
pure and complete samples based on several tens of observations.
Our situation is somewhat in between, as we generally have more
observations than \citet{sesar07} but fewer than W09 and S10, so we
need to thoroughly understand how contamination affects our
sample; in this we are aided by the fact that many of our candidate
RRLs have enough observations to enable a period to be estimated.
\par
Following \citet{vivas04} we estimated the period of the 265 RRL
candidates that have at least 12 observations in the $i$ band. To do
this we adopt the algorithm described by \citet{lafler65}, defining the
parameter:
\begin{equation}
\label{eq:theta}
\Theta = \frac{\sum\limits_i (m_i - m_{i+1})^2}{\sum\limits_i (m_i - \bar{m}_i)^2} \\
\end{equation}
where $m_i$ is the magnitude at phase $\phi_i$, $\bar{m}_i$ is the mean magnitude
and the phase $\phi$ is defined in the usual way as:
\begin{equation}
\label{eq:phase}
\phi = t / P - \lfloor t / P \rfloor
\end{equation}
where $P$ is the period and $\lfloor x \rfloor$ is the integer part of $x$.
In our implementation of the algorithm we considered an array of $10^7$ trial periods
in the $0.2-1~\mathrm{d}$ range and computed $\Theta$ for each of them.
The value of $P$ at which $\Theta$ is minimum is then taken as the best estimate
of the period.
\par
Since the number of observations for our RRL is relatively small, we
may expect that in some cases the period estimation is not
particularly accurate.
However, this is not a serious concern because
our aim here is simply to use periods to estimate our
contamination, not to use them in our distance estimates.
A number of our RRLs exist in the VSX database (see Section
\ref{sec:cross_match} and accompanying electronic table) and so,
for those that have a robust period, we compared our estimate
to the VSX value.
We found 88 such cases; only in 29 of them were our periods
close enough (defined as within $0.05\%$ of each other)
to be considered reliable.
The fact that only a minority of our periods can be
considered reliable is not a major problem as we do not use
them in our later analysis of the halo density profile but we
caution against their use in e.g. Period$-$Luminosity
relationships or other applications that require a precise
estimate of the period.  
\par
Another important concern is how significant the period thus computed
really is, as the above procedure will produce a minimum regardless of whether or not
a period is present.
\citet{lafler65} introduced the quantity:
\begin{equation}
\label{eq:lambda}
\Lambda = \frac{\Theta(\mathrm{At~incorrect~period})}{\Theta(\mathrm{At~true~period})}
\end{equation}
as a diagnostic for whether or not a period is indeed
present, i.e. objects with a significant period would have a higher
value of $\Lambda$.
\citet{lafler65} show that a minimum value
$\Lambda_{\mathrm{min}} = 4$ provides good discrimination between
light-curves which do have a period 
($\Lambda > \Lambda_{\mathrm{min}}$) and those which do not
($\Lambda < \Lambda_{\mathrm{min}}$) when the number of
observations is between $10-40$, which is the case with most of
our RRL candidates.
We therefore adopted $\Lambda_{\mathrm{min}} = 4$ to discriminate between
those objects which do have a period and those which do not. Note that when
the number of observations is larger $\Lambda_{\mathrm{min}}$ can be smaller:
\citet{saha90} argue for $\Lambda_{\mathrm{min}} = 3$ and
\citet{vivas04} use $\Lambda_{\mathrm{min}} = 2.5$.
We used the mean value of $\Theta$ from the whole range of trial
periods as our estimate of the numerator of Equation \ref{eq:lambda}. 
Out of 318 RRL candidates that passed the color and variability cut,
265 have at least 12 observations in the $i$ band. For these we
estimated a period using Equation \ref{eq:theta}.
In several cases the period initially computed was not good due to one or
two bad points, as revealed by visually inspecting the folded light-curve;
in these cases we recomputed the period and $\Lambda$ after removing those
points.
\par
We assumed that objects classified as RRLs either in the VSX database
or in the CDS database were real; therefore if one of them was flagged as
a false detection in the following steps it was not counted as such
in our final contamination estimate.
\par
As a first step on each period thus computed we imposed a cut
$\Lambda_{\mathrm{min}} = 4$ obtaining 203 periods.
This cut allowed us to estimate the contamination
due to variable objects not having a definite periodicity, or
with a period outside the $0.2-1~\mathrm{d}$ range and therefore
not RRLs.
This criterion yielded 62 objects that are expected not to be true
RRLs. We found that just 6 of these 62 are reported as RRLs by either
the VSX or the CDS database and we chose to consider them RRLs
too, meaning that this $\Lambda$ cut yields 56 false detections out of
265 candidate RRLs.
\par
This simple cut, while effective in removing a significant fraction of
interlopers is not sufficient: a simple way to see this is that after the cut
the number of RRL candidates with period $<0.4~\mathrm{d}$ (and therefore
probably RRLc) was about the same as the number of objects with period
$\approx 0.5-0.6~\mathrm{d}$ (and therefore probably RRLab), when in reality the
number of RRLc is much smaller than the number of RRLab. 
\par
As a second step we fitted all the 265 RRL candidates for which we
determined a period to the S10 templates and visually inspected all of them.
Objects that did not show an RRL type light-curves were flagged as ``interlopers"
regardless of the value of $\Lambda$.
It is worth noting, however, that the majority of these objects had
$\Lambda$ not much larger than 4 and so a slightly more restrictive cut
(such as $\Lambda>6$) would have been effective in removing the
majority of them; we chose nevertheless to flag them by visual
inspection because a few objects had large $\Lambda$ despite not
showing an RRL light-curve. It is also worth noting that the majority
of these objects had periods $<0.4~\mathrm{d}$, i.e. after removing
these interlopers the ratio of true RRLc to RRLab in our sample will
be reduced, bringing it into better agreement with previous studies.
We found 43 of these visually selected interlopers, 16 of which were
classified as RRLs by either the VSX or CDS and we chose to consider them as
RRLs too; so this second cut yielded additional 27 false detections out of
the initial 265.
\par
While we do not have sufficient data to characterize the nature
of these interloping objects, we may note that contamination from
non-stellar sources such as AGN should be negligible:
we cross-correlated our sample with the SDSS Quasar Catalog
Seventh Data Release \citep{schneider10} comprising 105,783 quasars
and did not find any match; this is consistent with the finding of
W09 who show that low redshift quasars in the \citet{bramich08}
data-set have mostly magnitudes in the
$20~\mathrm{mag}\lessapprox g^{\mathrm{SDSS}}\lessapprox 22~\mathrm{mag}$
range, well below our magnitude limit $i=19~\mathrm{mag}$.
We therefore conclude that most contaminating objects are probably
variable stars.
There are a number of potential suspects, including contact
binaries. These objects are troublesome because they can have periods
similar to RRL and, for poorly sampled light-curves, could be confused
with RRLc.
The relatively high number of cases in which an object passes
our variability cuts due to one or two points being far off the
mean \footnote{As discussed in Section \ref{sec:rrl},
these cases were dealt with by refitting each object after
eliminating each point in turn and excluding objects for which
one of the fits was good; this reduced the number of candidate
variables from 19556 to 14072 objects but a few interlopers may remain
in the catalog} suggests that an important class of interlopers may be
represented by detached eclipsing binary stars (EB) where the point(s)
far off the mean are observed at eclipse.
The light-curve of a detached EB is far too complex
for a period to be reliably found with the few observations we
have (we checked this using detached EBs data from the literature)
so the period finding procedure will produce meaningless results.
The interloping EBs, however, will be revealed by the
$\Lambda>4$ cut we used or by our visual inspection.
\par
One last concern that needs to be addressed
in more detail is a possible contamination by $\delta$
Scuti stars.
S10 note that the majority of false RRL candidates in \citet{sesar07}
turned out to be $\delta$ Scuti stars, in good agreement with
their predictions.
On the other hand $\delta$ Scuti stars contamination is
less problematic in W09, probably because, unlike \citet{sesar07}
they have enough points to estimate a period: their Figure 8
shows the period distribution for all their RRL candidates
and $\delta$ Scuti stars clearly stand out against the other
candidates as a peak at $P<0.1~\mathrm{d}$; these objects are
then removed from further consideration via appropriate cuts
\citep{watkins09}.
\par
Our procedure to address this issue is as follows:
\begin{enumerate}
\item
Taking all objects with $\Lambda > 4$, we select those which have a
bad fit to the S10 templates (defined as having a $\chi^2$ probability
less then 0.001). We found 59 such objects from an initial sample of 203.
\item
For these 59 objects with bad fits we tried to find periods
in the $0.05-0.2~\mathrm{d}$ range, appropriate for $\delta$
Scuti stars, and computed the $\Lambda$ parameter $\Lambda_{0.05-0.2}$.
\item
We compared $\Lambda$ to $\Lambda_{0.05-0.2}$ to decide
which of our candidate RRLs were likely $\delta$ Scuti stars.
\end{enumerate}
\par
The last step requires some care: the simplest idea is to
consider as $\delta$ Scuti all objects with
$\Lambda_{0.05-0.2}>\Lambda$, since in this case the period
found in the $0.05-0.2~\mathrm{d}$ range (typical of
$\delta$ Scuti stars) should be more significant than the one
in the $0.2-1~\mathrm{d}$ range (typical of RRLs).
This however is likely to produce an overestimate of the
contamination, as a visual inspection reveals that many
objects badly fitted by S10 templates have nevertheless
light-curves that closely resemble those of an RRL.
We therefore decided to keep those badly fitted objects
that show RRL-like light-curves as RRL and not include
them when computing the contamination, and adopt the
$\Lambda_{0.05-0.2}>\Lambda$ criterion only for those objects
that were not obviously RRLs.
Although some high-amplitude $\delta$ Scuti stars can have
light-curves similar to RRL and hence may not be rejected,
we assume that the number of these is negligible.
The criterion $\Lambda_{0.05-0.2}>\Lambda$ yields 23 objects
and only 5 of those, upon visual inspection, do look like
RRLs. A further 6 are reported as RRL by either the
VSX or the CDS database, so we considered the remaining 12 as likely
$\delta$ Scuti stars interlopers.
\par
Thus our contamination estimation yields 56 objects
with $\Lambda<4$, 27 visually selected interlopers and 12
$\delta$ Scuti candidates (in all cases objects found in the
VSX or CDS database are not included in this calculation). This
implies that there are a total of 95 non-RRLs in a
sample of 265 RRL candidates with a period estimate, so our
final estimate for the contamination fraction in our sample
is $95/265=0.36$; this factor will be assumed
when computing the space density of RRLs
in the halo (Section \ref{sec:solar}).
Our estimate leaves us with $265-95=170$ objects likely to be
bona fide RRLs among those with at least 12 observations;
extrapolating to the whole sample this translates to 204 RRLs. 
\par
\begin{figure}
\begin{center}
\plotone{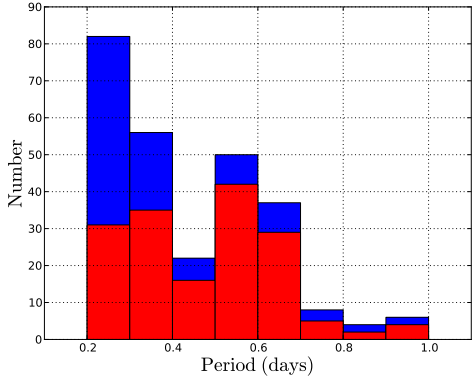}
\end{center}
\caption{Blue histogram: period distribution for 265 objects with at least 12 observations.
Red histogram: period distribution for 170 bona fide RRLs.}
\label{fig:per_hist}
\end{figure}

Figure \ref{fig:per_hist} shows the period distribution for all 265
objects with more than 12 epochs (blue histogram) and the
distribution for the 170 bona fide RRLs (red histogram).
Note that for most objects with $\Lambda<4$, the
tentative periods recovered by Equation \ref{eq:theta} cluster in the
$0.2-0.3~\mathrm{d}$ range; the red histogram
shows a smaller peak in the $0.2-0.4~\mathrm{d}$ range and
a larger peak in the $0.4-0.6~\mathrm{d}$ range, corresponding to
RRLc and RRLab respectively.
The complete list of the 265 periods and their $\Lambda$ (including
the epoch(s) removed for finding the period, if any, and their nature
as interlopers or $\delta$ Scuti)
is given in an accompanying electronic table.
\par
\begin{figure}
\begin{center}
\plotone{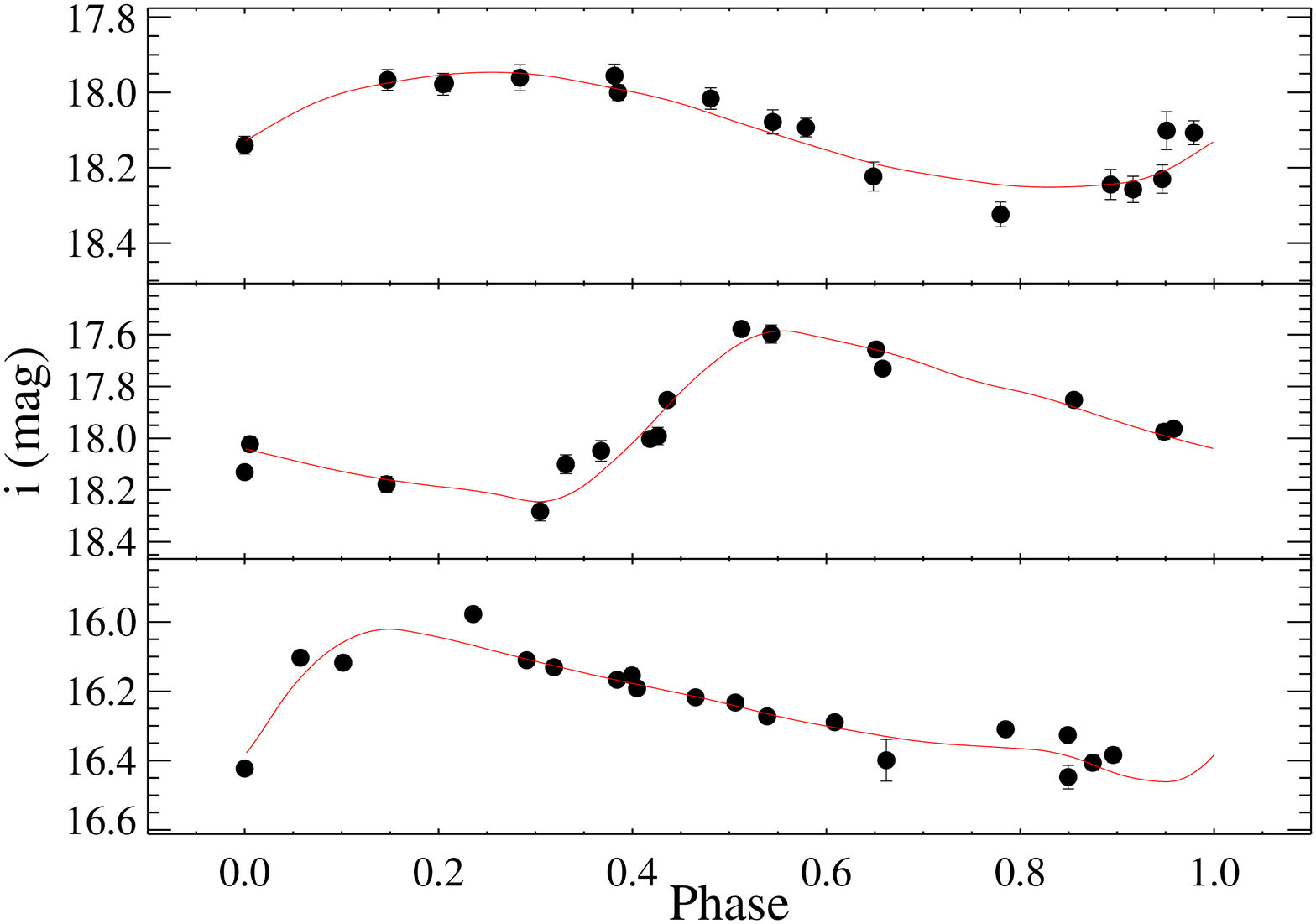}
\plotone{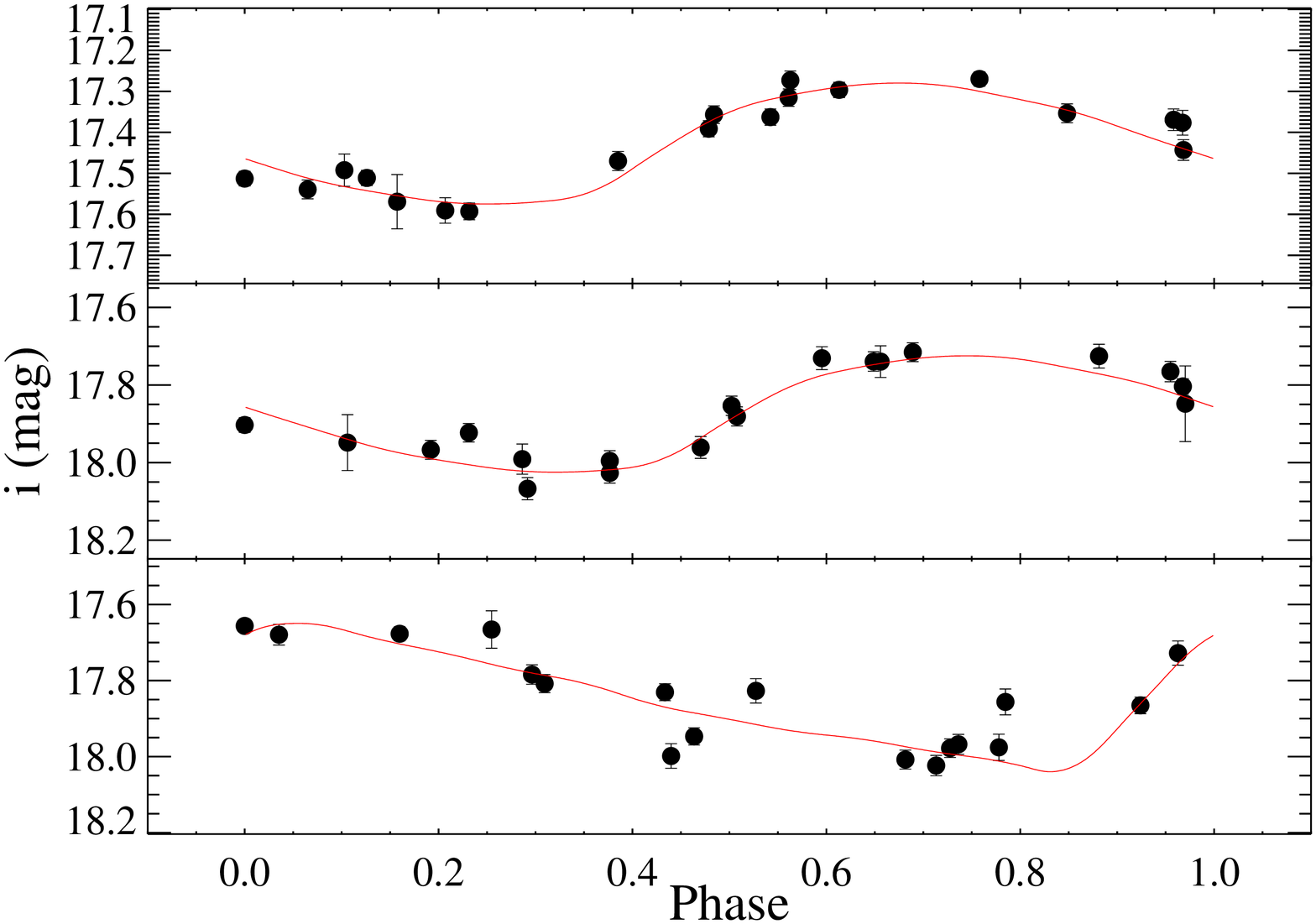}
\plotone{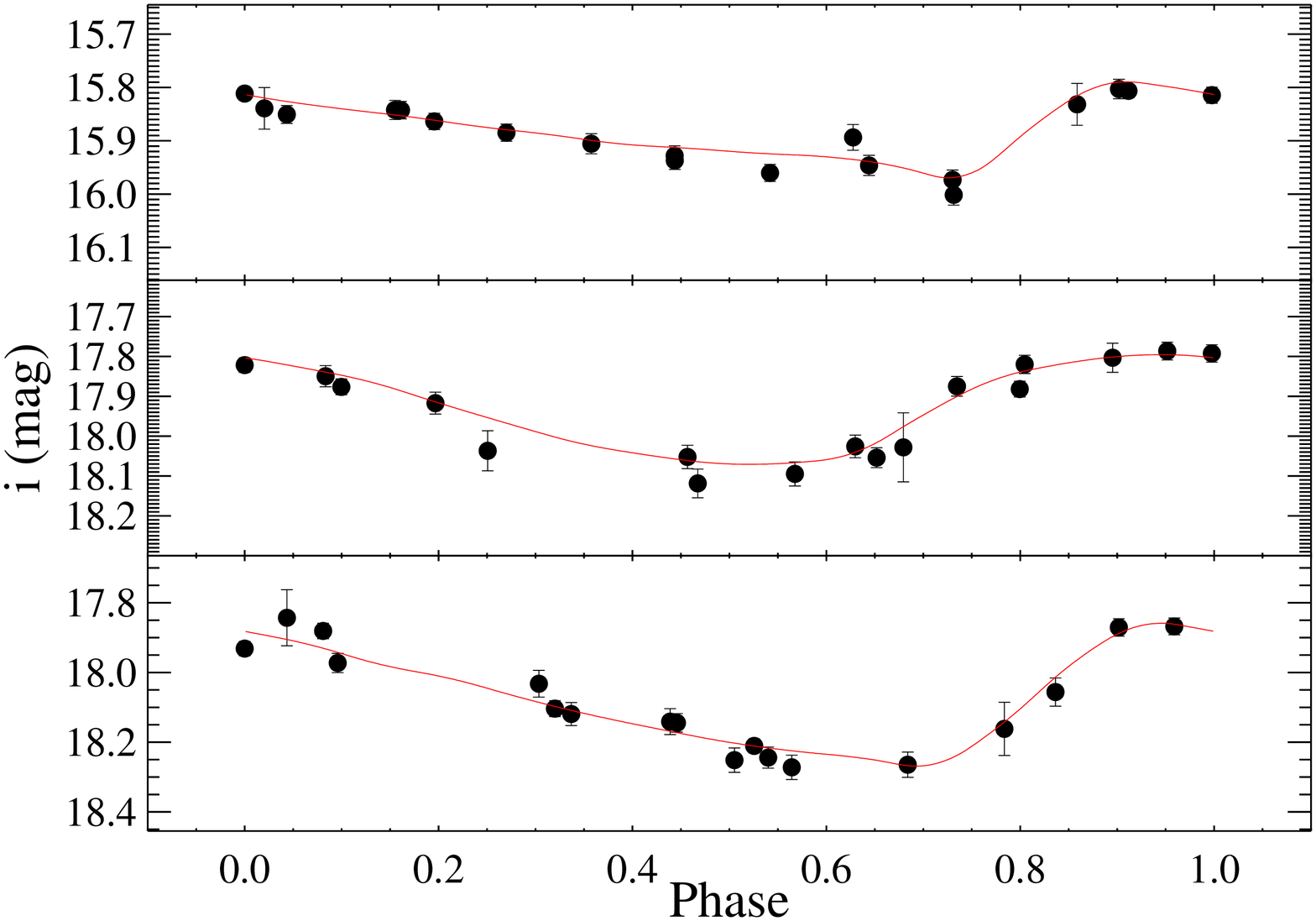}
\end{center}
\caption{Folded $i$ light-curves of selected RRL candidates.}
\label{fig:lc1}
\end{figure}

Figure \ref{fig:lc1} shows examples of folded light-curves and their fit to
the S10 templates; the complete light-curves are provided in
accompanying online files.
\section{Distance Estimation}
\label{sec:dist}
Although there are period-luminosity relationships for RRLs
\citep[e.g.][]{caceres08}, we choose not to use them for our study
because, as mentioned above, we do not have accurate periods for our
entire sample.
Therefore the best we can do is to attempt to estimate heliocentric
distance $d_\mathrm{h}$ from their mean unreddened $i_0$ 
magnitude alone
\citep[the reddening correction is performed via the maps of][]{schlegel98}
by comparing them to S10 unreddened absolute
magnitudes, which we derived using the data they provide.
We then calculate the S10 sample mean, which we use as our best estimate of 
$\langle I_0 \rangle$ for RRLs in the galactic halo.
After converting to the Xuyi system we find
$\langle I_0 \rangle = 0.57~\mathrm{mag}$, with a dispersion of 
$0.08~\mathrm{mag}$.
Note that by sampling the S10 absolute magnitude distribution we are
effectively marginalizing over the periods; moreover, as S10 assume a mean
halo metallicity [Fe/H]=-1.5 \citep[from][]{ivezic08} we are implicitly
doing the same.
Our distance indicator is thus given by the following equation:
\begin{equation}
\label{eq:dist}
d_\mathrm{h} = 10 ^ {(i_0 - \langle I_0 \rangle  + 5) / 5)} / 1000~\kpc.
\end{equation}
\par
\begin{figure}
\begin{center}
\plotone{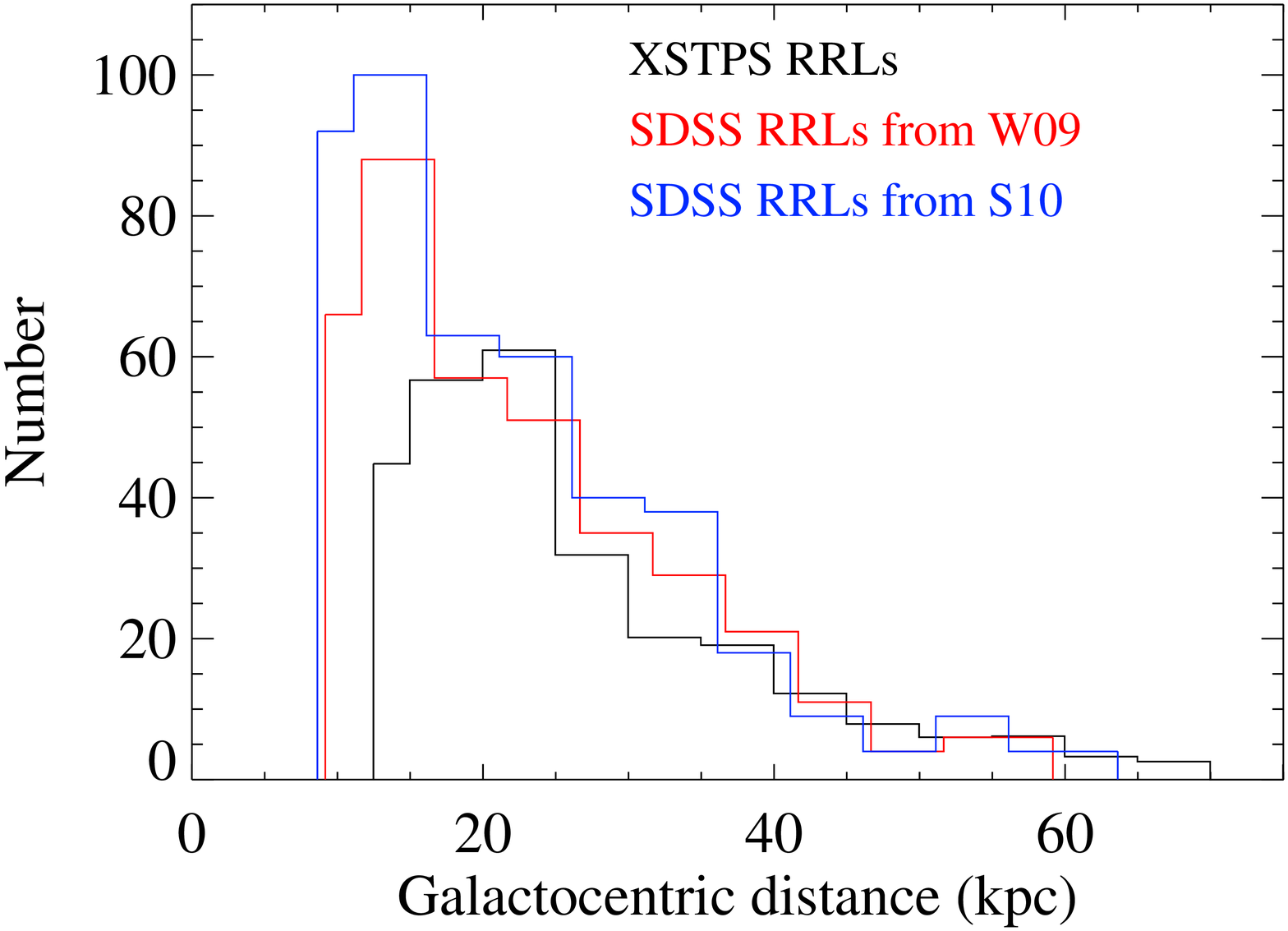}
\end{center}
\caption{Histogram of the galacto-centric distances for the \xdss, W09, and S10
RRL samples; the \xdss~data are corrected for efficiency and
contamination;
only W09 and S10 RRLs with galacto-centric distances $<65~\kpc$ are shown.}
\label{fig:galacto}
\end{figure}

We then construct an histogram of distances for our sample by
correcting both for efficiency $\epsilon$ (described
in Section \ref{sec:eff}) and for contamination $c$ (described in
Section \ref{sec:period}). We do this by assigning each RRL candidate
a weight $w = (1-c) / \epsilon$ when constructing the histogram,
with $c=0.36$; note that while $\epsilon$ varies from object to
object, $c$ is the same for the whole sample.
The result is shown in Figure \ref{fig:galacto}, which also shows for comparison
the histograms for the W09 and S10 samples, where, for these, we
only included objects up to a galacto-centric distance $65~\kpc$
and did not perform any correction for efficiency or contamination
(W09 explicitly state that $\epsilon\approx 1$ for their sample).
The figure shows an excess RRL at $\approx 20~\kpc$ for the \xdss~sample
and a smaller one at $\approx 35~\kpc$ for the W09 and S10 samples, in both
cases exactly where the Sagittarius Stream is, so these excesses
are probably due to contamination from stream stars;
this is consistent with W09 who find that 55 RRLs in their sample are
associated with the stream.
%
%

We estimated our distance uncertainties using a MC process.
For each candidate RRL we randomly sampled the distribution
of absolute $I_0$ magnitudes from S10 (converted to the Xuyi
photometric system). We did this 10,000 times and
recomputed distances via Equation \ref{eq:dist}.
The standard deviation of these resamples is then taken to represent
the uncertainty in the distance.
Typical MC uncertainties in the heliocentric distances are
about $4\%$, which is entirely consistent with the expected
uncertainties associated with the spread in $I_0$ 
(the relative uncertainty in distance $d$ is given
by $\delta d/d = 0.2\log(10)\delta\langle I_0\rangle=0.037$).

Since period-luminosity relationships are usually
metallicity-dependent and we do not have metallicity information for
our sample, this will add an uncertainty of 0.07 mag in $V$ (see
section 4.1 of S10 for details of how this value is determined).
Furthermore, as discussed in \citet{vivas06}, the change in absolute
magnitude with the RRL evolution off the Zero Age Horizontal Branch
will affect the distance estimation, introducing an additional
dispersion of $0.08~\mathrm{mag}$ in $V$.
Even though both of these contributions should already be taken into
account when we sample the S10 absolute magnitude distribution (since
this will include RRL of various metallicities, consistent with the
metallicity spread of the halo, and RRL in a variety of evolutionary
states), we choose to be conservative and include them in our total
error budget.
Assuming that the spread in $V$ is similar to the spread in our Xuyi
$i$-band for these two terms, this means that the total uncertainty in
absolute magnitude is
$\delta\langle I_0\rangle\approx\sqrt{0.08^2+0.07^2+0.08^2} = 0.13$. 
When combined with the photometric uncertainties in $i_0$, which are
typically around a few hundreths of a magnitude for our Xuyi sample,
we find our distances are accurate to $\approx 6-7\%$, consistent
with similar values found by S10 and \citet{vivas06}.
The distances and the uncertainties from the MC procedure
are given in an accompanying electronic table.

\begin{figure}
\begin{center}
\plotone{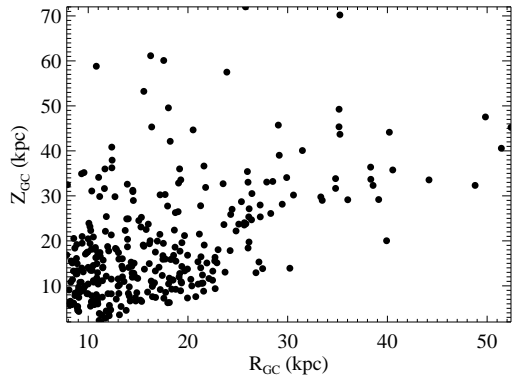}
\end{center}
\caption{RRL distribution in the 
$R_\mathrm{GC}\equiv\sqrt{X_\mathrm{GC}^2+Y_\mathrm{GC}^2},
Z_\mathrm{GC}$ plane.}
\label{fig:rgc-zgc}
\end{figure}

It is now easy to derive the spatial distribution of RRLs in
a galacto-centric reference frame
$X_\mathrm{GC}, Y_\mathrm{GC}, Z_\mathrm{GC}$, which is shown
in Figure \ref{fig:rgc-zgc} in the
$R_\mathrm{GC}\equiv\sqrt{X_\mathrm{GC}^2+Y_\mathrm{GC}^2},
Z_\mathrm{GC}$ plane; we assume a distance to the
Galactic Center ($\mathrm{GC}$) of $8~\kpc$ and $Z_\mathrm{GC}$
is directed toward the North Galactic Pole; the figure shows
all the RRLs in the sample, including the few that are beyond our
magnitude limit $i=19~\mathrm{mag}$
\section{The halo density profile}
\label{sec:solar}
\subsection{Spherical Halo Models}
\label{subsec:spherical}
We now use our data to study the halo density profile in the
NGC and compare the results to those obtained using the W09 and S10 samples
for the SGC.
Note that the W09 and S10 samples have been constructed using the same SDSS
data and hence are not independent samples.
We start by considering spherically symmetric models in the form
$n=n(r)$ where $r$ is the galacto-centric distance.
\par
The halo density profile out to the outer halo has often been parameterized as a
double broken power law \citep[see W09 for a discussion and, for a
collection of recent results,][]{akhter12}; in particular
W09 fit a double power law to their data-set in the form:
\begin{equation}
\label{eq:density}
n(r) = n_0\begin{cases}
\left(\frac{R_0}{r}\right)^\alpha & \text{if $R_\mathrm{min}<r<R_0$,}
\\
\left(\frac{R_0}{r}\right)^\beta  & \text{if $R_0<r<R_\mathrm{max}$}
\end{cases}
\end{equation}
and derive the following values for the halo parameters 
in the SGC:
$(n_0, R_0, \alpha, \beta) = (0.26~\kpc^{-3}, 23~\kpc, 2.4, 4.5)$
\footnote{The value of the normalization has been misreported in W09 and should be
0.26 $\kpc^{-3}$, rather than the value of $2.6~\kpc^{-3}$ quoted in their paper
(Laura Watkins, private communication)}.
We fit Equation \ref{eq:density} to the \xdss, W09, and
S10 data-sets.
To fit Equation \ref{eq:density} we need to determine
the completeness limit of the \xdss, W09, and S10 data-sets
and use this limit as $R_\mathrm{max}$.
In Section \ref{sec:xdss} we used Figure \ref{fig:rms_mag} to argue that
the completeness limit for finding RRLs in \xdss~is around
$i = 19~\mathrm{mag}$; beyond that limit the photometric uncertainties
become too large to reliably identify RRLs by the variability cuts we
employed. Assuming a mean absolute magnitude $\langle I_0 \rangle =
0.57~\mathrm{mag}$ 
(see Section \ref{sec:dist}) we derive via Equation \ref{eq:dist} a maximum
distance for our sample of $48.62~\kpc$; we therefore adopt
a maximum distance, $R_\mathrm{max} = 49~\kpc$ that allows us to sample the
halo well past the broken power law regime (the break takes place at
$R_0 \approx 23~\kpc$, see W09).
We use this same value of $R_\mathrm{max}$ for the W09 and S10 data
for three reasons: first for consistency with our data, second because at
higher distances the number of RRLs becomes so low that the quality of the fit
may be affected (unlike W09 we fit for the density, not its volume integral),
and finally because at higher distances the W09 and S10 samples are affected by
local over-densities that may negatively impact a fit to a smooth profile
(again this may be less of a concern for W09 who fit for the volume
integral of the density); we also remove from the fit the few
RRLs in all samples with distance $r<9~\kpc$.
We compute the measured $n(r)$ to be fit to Equation \ref{eq:density}
by binning the RRLs in $1~\mathrm{kpc}$ bins in the
$9-49~\mathrm{kpc}$ range, assigning each candidate a weight 
$w=(1 - 0.36) / \epsilon$ where $0.36$ is the contamination computed
in Section \ref{sec:period} and $\epsilon$ is the efficiency
computed in Section \ref{sec:eff}, and dividing by the bin volume.
Note that the bin volumes, being centered on the GC and not on the
Sun, cannot be naively computed as $V=\frac{\Omega}{3}(R_2^3 - R_1^3)$
(where $\Omega = 376.75~\deg^2$ is the \xdss~stripe area and $R_1$, $R_2$
are the bin boundaries) because the stripe area is valid for a helio-centric
coordinates system.
Therefore we computed them by numerical integration, generating random RRLs
distributed in a large cubic volume around the GC and counting the fraction
of these RRLs that would both fall within each radial bin and that would be
visible in our stripe; this fraction, multiplied by the cubic volume,
gives the bin volume; this is very important for RRLs whose
galacto-centric distance is not much greater than our adopted Sun-GC
distance ($8~\kpc$) but becomes less and less important for RRLs at
greater distances. Uncertainties are estimated using Poisson
statistics, dividing the square root of the weighted number of RRL
candidates in each $1~\mathrm{kpc}$ bin by the bin volume.
We do the same for the W09 and S10 data-sets, except that in that
case we assign $w=1$ to each RRL
since both samples are free of contamination and have
$\epsilon\approx 1$ \citep{watkins09, sesar10} due to their good
light-curve sampling.
\par
The fit results are summarized in Table \ref{tab:fit}.
\begin{table}
\caption{Summary of fits to double power law (Equation 
\ref{eq:density}).}
\begin{tabular}{c c c c c c}
\hline
Data & $n_0~(\kpc^{-3})$ & $R_0~(\kpc)$ & $\alpha$ & $\beta$ & $\chi^2/\mathrm{dof}$ \\
\hline
\xdss & $0.42 \pm 0.16$ & $21.5 \pm 2.2$ & $2.3 \pm 0.5$ & $4.8 \pm 0.5$ & $0.65$ \\
\xdss $^{a}$ & $0.40 \pm 0.19$ & $21.5 \pm 2.6$ & $2.4 \pm 0.5$ & $4.7 \pm 0.5$ & $0.64$ \\
W09   & $0.28 \pm 0.11$ & $22.9 \pm 2.5$ & $2.8 \pm 0.3$ & $4.9 \pm 0.5$ & $1.22$ \\
S10   & $0.21 \pm 0.08$ & $25.5 \pm 2.4$ & $2.9 \pm 0.3$ & $5.2 \pm 0.6$ & $1.05$ \\
\hline
\end{tabular}
\tablecomments{$^a$ The \xdss~RRLs at $\approx 20~\kpc$ have been
removed from this fit to account for possible contamination from
Sagittarius Stream stars.
}
\label{tab:fit}
\end{table}
This table shows that the three data-sets give results
consistent with each other at the $\approx 1\sigma$ level
(albeit with a rather large formal uncertainty)
for all the halo profile parameters.

One concern that should be addressed is a possible contamination
of our data by
Sagittarius Stream stars in the \xdss~stripe area
(see for example Figure 1 of \citealt{belokurov06}).
\citet{law10} present a model of the Sagittarius Stream based on
a triaxial Milky Way halo which reproduces most observational
constraints and present the results of their $N-$body simulations in
the form of three dimensional positions and velocities of 10,000 Sagittarius
Stream particles both for the leading and the trailing arm
of the stream; their results show that the distribution of 
particles is strongly peaked at $\approx 20~\kpc$ in the area covered
by \xdss~data.
Therefore we redid the fit removing the data with $19~\kpc<r<21~\kpc$.
Results of this fit are given in the second row of Table \ref{tab:fit}
and, reassuringly, show very little variation from the original fit.
\par
The W09 and S10 samples are also contaminated by Sagittarius Stream
stars (W09).
The \citet{law10} model shows that at the SDSS Stripe 82
location the star distribution peaks at $\approx 20$ and $40~\kpc$
but these peaks are much less sharp than the one exhibited by stream
stars at the ~\xdss~location; we checked that removing RRLs at those
distances in the W09 and S10 samples does not affect the fit much so we
did not correct for Sagittarius Stream contamination for the W09 and S10
samples. More detailed studies, for example folding in the velocities
or chemistry of the stars, would be able to address the issue of
Sagittarius contamination more robustly, but we do not wish to attempt
such analyses here.
\par
\begin{figure}
\begin{center}
\plotone{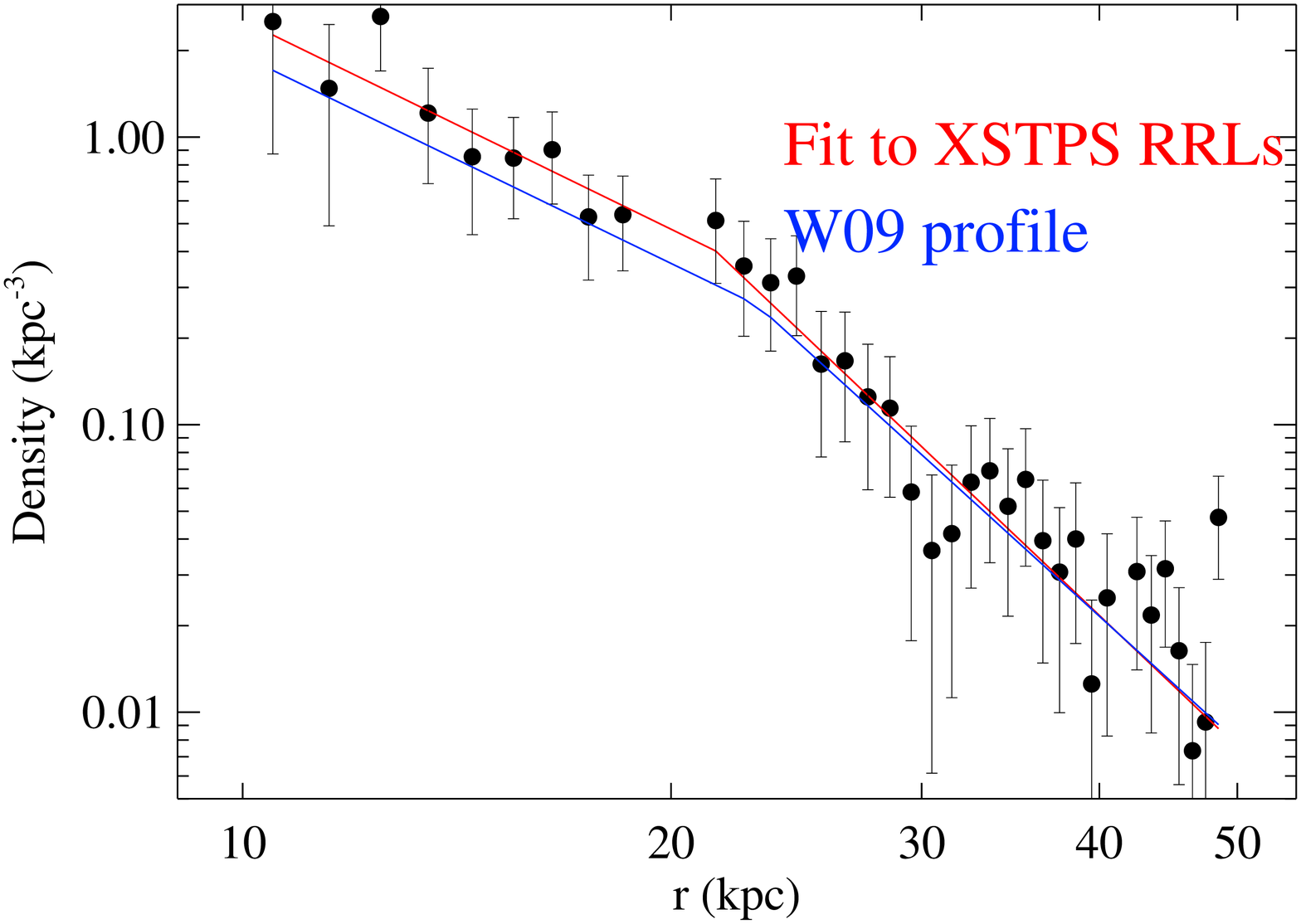}
\end{center}
\caption{RRL number density as a function of galacto-centric distance
in the $9-49~\kpc$ range with the \xdss~RRLs at $\approx 20~\kpc$ removed from
the fit to account for possible contamination from Sagittarius Stream stars.
The red line shows the fit to a double power law given by
Equation \ref{eq:density}.
The blue line shows the W09 profile with $n_0=0.26~\kpc^{-3}$.}
\label{fig:fit_nosag}
\end{figure}
\par

Our halo profile fit is illustrated in Figure \ref{fig:fit_nosag},
which also shows the W09 profile with $n_0=0.26~\kpc^{-3}$.
The agreement between the \xdss~and W09 sample is excellent,
whereas from Table \ref{tab:fit} we can see that the S10 sample is
somewhat more discrepant; in all cases however we find evidence for a
steepening of the halo density profile beyond $23-25~\kpc$, consistent
with both W09 and S10.
\par
Therefore we may conclude that the halo density profile in the NGC in
the $9-49~\kpc$ range (as sampled by the RRLs in the \xdss~data-set) is in
broad agreement with the density profile in the SGC in the same range
(as sampled by the RRLs in the W09 and S10 data-sets).
Both NGC and SGC data can be adequately described by a double
power law given by Equation \ref{eq:density} with parameters
consistent across all three data-sets at the $1\sigma$ level.
\subsection{Non-spherical Halo Models}
\label{subsec:oblate}
It is interesting to check whether our RRLs, as well as the W09 and
S10 ones, can constrain non-spherical models of the halo as there is
evidence that the halo density profile can be best described by
oblate or triaxial models.
\par
\citet[][hereafter D11]{deason11} use a maximum likelihood 
method to study the  density profile of BHB and blue straggler
stars from SDSS Data Release 8; their sample comprises
$\approx 20,000$ stars over a $14,000~\deg^2$ area encompassing both
the Northern and the Southern Galactic hemispheres;
their main finding is that oblate and triaxial
halo models describe their data better than spherical ones, with
an oblate double power law model giving the highest likelihood
overall.
\citet[][hereafter S11]{sesar11b} use Canada-France-Hawaii Legacy
Survey\footnote{\url{http://www.cfht.hawaii.edu/Science/CFHTLS/}} data in a
$170~\deg^2$ area along four lines of sight to
probe the galactic halo via the distribution of near-turnoff main-sequence
stars up to an heliocentric distance of $\approx 35~\kpc$.
S11 too find that the halo profile can best be described by an oblate
double power law as well as detecting both the Sagittarius and the
Monoceros Streams.
Non-spherical halo models have also been considered by \citet{vivas06}
who find that non-spherical single power law models better describe
their RRL sample.
\par
\begin{figure}
\begin{center}
\plotone{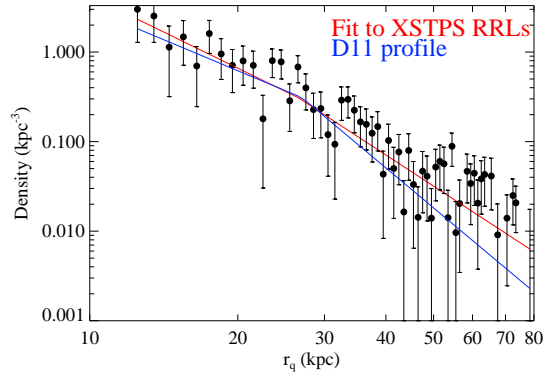}
\end{center}
\caption{RRL number density as a function of $r_q$
in the galacto-centric distance $r=9-49~\kpc$ range
with the \xdss~RRLs at $\approx 20~\kpc$ removed from
the fit to account for possible contamination from Sagittarius Stream stars.
The red line shows the fit to a double
power law given by Equation \ref{eq:oblate} with $q=0.59$.
The blue line shows the D11 profile with $n_0=0.31~\kpc^{-3}$ from our fit
(see Table \ref{tab:oblate}).}
\label{fig:oblate}
\end{figure}

We reanalyzed the \xdss, W09, and S10 data-sets in light of the
D11 and S11 results: we started by
setting up a cylindrical coordinate system with origin in the
Galactic Center, described by
$R_\mathrm{GC}\equiv\sqrt{X_\mathrm{GC}^2+Y_\mathrm{GC}^2}$,
$\phi\equiv\arctan(Y_\mathrm{GC}/X_\mathrm{GC})$, and
$Z_\mathrm{GC}$ where
$X_\mathrm{GC}, Y_\mathrm{GC}, Z_\mathrm{GC}$ are 
galacto-centric coordinates and consider a non-spherical double
power law model:
\begin{equation}
\label{eq:oblate}
n(r_q) = n_0\begin{cases}
\left(\frac{R_0}{r_q}\right)^\alpha & \text{if $R_\mathrm{min}<r_q<R_0$},
\\
\left(\frac{R_0}{r_q}\right)^\beta  & \text{if $R_0<r_q<R_\mathrm{max}$}
\end{cases}
\end{equation}
where $r_q\equiv\sqrt{\rgc^2 + \zgc^2q^{-2}}$ and $q$ is a new
parameter describing the flattening of the halo: $q<1$ for
oblate models, $q>1$ for prolate models and $q=1$ for spherical
ones, in which case Equation \ref{eq:oblate} reduces to
Equation \ref{eq:density}; the meaning of the other symbols is
the same as Equation \ref{eq:density}.
D11 find that a model described by Equation 
\ref{eq:oblate} with parameters
$R_0=27\pm 1~\kpc$, $\alpha=2.3\pm 0.1$, 
$\beta=4.6^{+0.2}_{-0.1}$, and
$q=0.59^{+0.02}_{-0.03}$, best fit their data; note that
since they are analyzing different stellar tracers (BHBs and blue
stragglers, rather than RRLs), we are not concerned with their value
for the normalization.
First we considered the D11 results, fitting Equation \ref{eq:oblate}
assuming $q=0.59$; as usual we considered RRLs with
a galacto-centric distance $9~\kpc<r<49~\kpc$ for all three data-sets
and binned them in $1~\kpc$ bins as we did for the spherical
cases; we also removed again RRLs with $19~\kpc<r<21~\kpc$ in the
\xdss~data-set to avoid contamination from stars in
the Sagittarius Stream.
Figure \ref{fig:oblate} shows the density profile $n(r_q)$ for
the \xdss~data and the fit to Equation \ref{eq:oblate} with the
parameters given by Table \ref{tab:oblate};
also shown is the D11 profile with $n_0=0.38~\kpc^{-3}$ from our
fit.
Note that, even though we cut at $r=49~\kpc$, $r_q$ can reach
much higher values since the $Z$ component of $r_q$ is divided
by $q^2=0.348$
%
\begin{table}
\caption{Summary of fits to oblate double power law 
(Equation \ref{eq:oblate} with $q=0.59$)
with the \xdss~RRLs at $\approx 20~\kpc$ removed from
the fit to account for possible contamination from Sagittarius Stream stars.
The D11 results are taken from \citet{deason11}.}
\begin{tabular}{c c c c c c}
\hline
Data & $n_0~(\kpc^{-3})$ & $R_0~(\kpc)$ & $\alpha$ & $\beta$ & $\chi^2/\mathrm{dof}$\\
\hline
\xdss & $0.31 \pm 0.33$ & $26.5 \pm 8.9$ & $2.7 \pm 0.6$ & $3.6 \pm 0.4$ & $1.04$ \\
W09   & $0.43 \pm 0.16$ & $26.9 \pm 3.1$ & $2.1 \pm 0.3$ & $4.0 \pm 0.3$ & $0.69$ \\
S10   & $0.42 \pm 0.40$ & $26.2 \pm 7.4$ & $3.0 \pm 0.3$ & $3.8 \pm 0.3$ & $1.52$ \\
D11   & N.A.            & $27\pm 1$      & $2.3 \pm 0.1$ & $4.6^{+0.2}_{-0.1}$ & N.A.\\
\hline
\end{tabular}
\label{tab:oblate}
\end{table}
\par
Table \ref{tab:oblate} shows that agreement with the D11 result is excellent
for all three RRL data-sets for the break radius $R_0$ and at about
$1\sigma$ level for $\alpha$, while there is some tension (at
about $2\sigma$ level) for $\beta$ with the RRL data preferring lower
values.
It should be noted however that the formal errors from the fits are 
considerable (for example, for the \xdss~ and S10 samples the error on
$n_0$ is of the same magnitude as $n_0$ itself)
and the $\chi^2/\mathrm{dof}$ is rather large for the S10 sample,
suggesting that these current RRL data-sets are unable to provide strong
constraints on this model.
Of the three RRL samples we considered, W09 is both the one that
is better described by an oblate halo model with $q=0.59$ and the one for
which the agreement with the D11 result is best.
\par
We then considered the S11 results, which differ quite considerably
from those of D11 (this may be due to the fact that the two
groups use different tracer populations, with D11 using BHB and Blue
Straggler stars while S11 use near-turnoff main-sequence stars).
While the agreement for the break radius $R_0$ is excellent, there is
tension at the few $\sigma$ level 
for $\alpha$ and $\beta$; S11 also find evidence for a less oblate
halo ($q=0.70\pm 0.01$).
\begin{table}
\caption{Summary of fits to oblate double power law 
(Equation \ref{eq:oblate} with $q=0.7$)
with the \xdss~RRLs at $\approx 20~\kpc$ removed from
the fit to account for possible contamination from Sagittarius Stream stars.
The S11 results are taken from \citet{sesar11b}; the $n_0$ parameter refers
to near-turnoff main-sequence stars so it cannot be compared to the RRL
values.}
\begin{tabular}{c c c c c c}
\hline
Data & $n_0~(\kpc^{-3})$ & $R_0~(\kpc)$ & $\alpha$ & $\beta$ & $\chi^2/\mathrm{dof}$\\
\hline
\xdss & $0.21 \pm 0.14$ & $28.5 \pm 5.6$ & $2.8 \pm 0.4$ & $4.4 \pm 0.7$ & $0.8$ \\
W09   & $0.28 \pm 0.12$ & $27.6 \pm 3.3$ & $2.5 \pm 0.3$ & $4.3 \pm 0.4$ & $1.1$ \\
S10   & $0.15 \pm 0.04$ & $34.6 \pm 2.8$ & $2.8 \pm 0.2$ & $5.8 \pm 0.9$ & $1.1$ \\
S11   & $1450 \pm 50$   & $27.8 \pm 0.8$ & $2.62\pm 0.04$ & $3.8\pm 0.1$ & $3.9$ \\
\hline
\end{tabular}
\label{tab:oblate_sesar}
\end{table}
\begin{figure}
\begin{center}
\plotone{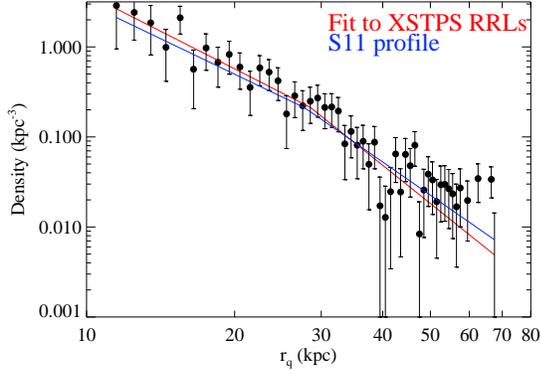}
\end{center}
\caption{RRL number density as a function of $r_q$
in the galacto-centric distance $r=9-49~\kpc$ range
with the \xdss~RRLs at $\approx 20~\kpc$ removed from
the fit to account for possible contamination from Sagittarius Stream stars.
The red line shows the fit to a double
power law given by Equation \ref{eq:oblate} with $q=0.7$.
The blue line shows the S11 profile with $n_0=0.32~\kpc^{-3}$ from our fit
(see Table \ref{tab:oblate_sesar}).}
\label{fig:oblate_sesar}
\end{figure}
\par
Our results using this S11 flattening are presented in Figure
\ref{fig:oblate_sesar}, which shows the density profile $n(r_q)$ for
the \xdss~data and the fit to Equation \ref{eq:oblate}, with the
parameters given in Table \ref{tab:oblate_sesar}; also shown is the
S11 profile normalized to match our fit.
Table \ref{tab:oblate_sesar} shows good agreement in $R_0$, $\alpha$ and
$\beta$ between \xdss, W09 and S11, although there is some tension with
S10. It should be noted that again the formal uncertainties are rather large.
The $\chi^2/\mathrm{dof}$ are somewhat lower than the $q=0.59$ case
for \xdss~and S10 (but higher for W09), suggesting that current RRL
data prefer a less oblate halo than D11 and agree more with S11.
It should be emphasized however that the fact that the three RRL samples
provide a decent fit to both spherical and oblate halo models shows that they
are not able to provide strong constraints for non-spherical halo
models and that larger samples with greater sky coverage are
needed.
Wide-field surveys are beginning to make this possible, for example
the work of \citet{sesar13} who use RRL from LINEAR to measure a halo
flattening of $0.63\pm0.05$.
\par
We finally note that \citet{akhter12} present a compilation of
recent studies that modelled the halo as a double power law
using both a variety of data-sets and a variety of tracers and
considering both spherical and non-spherical models.
These studies find a great variety of values for the break radius,
ranging from $23$ kpc (W09) to $45$ kpc \citep{keller08};
the inner power law index $\alpha$ is much better constrained,
with values mostly in the $2.3-2.7$ range; the outer power
law index $\beta$ is less well constrained with values ranging
from $3.6$ to $5$; of all these studies we find the best
agreement with W09 but clearly more work is needed as the
spread in values found by the different studies, especially
for the break radius, is still very large.
\section{Conclusions}
\label{sec:conc}
We have presented a sample of 318 candidates RRLs observed by the \xdss~in
an $\approx 376.75~\deg^2$ stripe in the NGC, selected via a
combination of SDSS color cuts and variability information afforded by
the multi-epoch nature of these \xdss~observations.
We have quantified the efficiency of our discovery procedure via Monte
Carlo methods and, by estimating the period for those candidate RRLs
that have enough observations, we have also quantified the effect of
contamination from non-RRL passing our color and variability cuts.
We have estimated a distance to the RRLs by assuming a mean
absolute magnitude
and have used this information to probe the halo density
profile in the NGC in the $9-49~\kpc$ range.
We found that the halo can be well approximated by a double power law,
as found in the SGC by W09.
There is agreement at the $1\sigma$ level between our model and the
models derived for the SGC using the W09 and S10 data-sets, and,
after removing RRLs at $r\approx 20~\kpc$ in our sample from the fit
to account for possible contamination from RRLs in the Sagittarius Stream,
the agreement between our sample and the SGC samples further
improves; in particular the agreement with the W09 result is excellent.
We considered non-spherical double power law models of the halo
density profile and found again agreement between the results
from our sample, the W09 and S10, and two samples using different halo tracers:
the D11 BHB and Blue Stragglers sample and the S11
near-turnoff main-sequence stars.
Our fits show that the \xdss~and S10 samples favor the less oblate
halo model of S11, whereas the W09 samples agree with the more oblate
model of D11. However, the fits are affected by large formal
uncertainties and, when comparing to D11, mediocre goodness of fit (at
least for the S10 sample).
Furthermore there could be population effects since these different studies
use a variety of halo tracer populations.
We therefore conclude that these RRL samples are not able to provide strong
constraints for non-spherical halo models and that larger samples with greater
sky coverage are needed.
This is already being made possible by projects such as LINEAR
\citep{sesar13}, the Catalina Survey \citep{drake13a}, the Palomar
Transient Factory \citep{rau2009} and Pan-STARRS \citep{kai2002}.
In the near future Gaia, ESA's space astrometry mission, will map the
entire sky repeatedly down to 20th magnitude \citep{debru12}, becoming a
valuable resource for RRLs and other standard candles
\citep{bono03b,eyer12}. Although most of the sky will be monitored
with an average of 70 epochs, certain regions will only have a few
tens of measurements \citep[see figure 3 of][]{debru12}. However, as
we have shown in this paper, even with a limited number of epochs it
is still possible to catalog and analyze the properties of RRLs. The
coming decade will provide unprecedented samples of RRLs, which will
be an invaluable resource for understanding the profile of our stellar
halo and illuminating the nature of the dark matter distribution
around our galaxy.
\section*{Acknowledgements}
We thank Shude Mao for suggestions, Laura Watkins for answering our questions
about her paper, Eric Peng and Brian Yanny for suggesting quality cuts for
querying the SDSS database, the SDSS help desk for help with accessing the
database itself, and the referee for a helpful and constructive report.
L.F. and M.C.S. acknowledge financial support from the Peking
University and CAS One Hundred Talent Funds, NSFC Grants 11173002 and 11333003.
H.B.Y, H.H.Z and X.W.L are partially supported by NSFC Grant 11078006.
H.B.Z. is partially supported by NSFC Grants 11078006, 10933004,
and 11273067, and the Foundation of Minor Planets of Purple
Mountain Observatory. 
This work was also supported by the following grants: the Gaia Research for European Astronomy Training (GREAT-ITN) Marie Curie network, funded through the European Union Seventh Framework Programme (FP7/2007-2013) under grant agreement no 264895; the Strategic Priority Research Program "The Emergence of Cosmological Structures" of the Chinese Academy of Sciences, Grant No. XDB09000000; and the National Key Basic Research Program of China 2014CB845700.

\label{lastpage}


\begin{thebibliography}{}

\bibitem[\protect\citeauthoryear{Akhter et al.}{2012}]{akhter12} 
Akhter S., Da Costa G.~S., Keller S.~C., Schmidt B.~P., 2012, ApJ, 756, 23

\bibitem[\protect\citeauthoryear{Alcock et al.}{1993}]{alcock93} 
Alcock C., et al., 1993, Natur, 365, 621

\bibitem[\protect\citeauthoryear{Alcock et al.}{1998}]{alcock98} 
Alcock C., et al., 1998, ApJ, 492, 190

\bibitem[\protect\citeauthoryear{Aubourg et 
al.}{1993}]{aubourg93} Aubourg E., et al., 1993, Natur, 365, 623

\bibitem[\protect\citeauthoryear{Belokurov et 
al.}{2006}]{belokurov06} Belokurov V., et al., 2006, ApJ, 642, L137 

\bibitem[\protect\citeauthoryear{Bond et al.}{2010}]{bond10} 
Bond N.~A., et al., 2010, ApJ, 716, 1 

\bibitem[\protect\citeauthoryear{Bono}{2003a}]{bono03a} Bono G., 
2003a, in Alloin D., Gieren W., eds, Lecture Notes in Physics,
Vol. 635, Stellar Candles for the Extragalactic Distance Scale, p. 85

\bibitem[\protect\citeauthoryear{Bono}{2003b}]{bono03b} Bono G., 
2003b, in Munari U., ed, ASP Conference Proceedings, Vol. 298, 
GAIA Spectroscopy: Science and Technology, p. 245

\bibitem[\protect\citeauthoryear{Bowell et al.}{1995}]{bowell95} 
Bowell E., Koehn B.~W., Howell S.~B., Hoffman M., Muinonen K., 1995, DPS, 
27, 1057 

\bibitem[\protect\citeauthoryear{Bramich et 
al.}{2008}]{bramich08} Bramich D.~M., et al., 2008, MNRAS, 386, 
887 

\bibitem[\protect\citeauthoryear{de Bruijne}{2012}]{debru12} de
  Bruijne J.H.J., 2012, Astrophys Space Sci, 341, 31

\bibitem[\protect\citeauthoryear{C\'aceres \& Catelan}{2008}]{caceres08}
C\'aceres C., Catelan M., 2008, ApJS, 179, 242

\bibitem[\protect\citeauthoryear{Deason, Belokurov, 
\& Evans}{2011}]{deason11} Deason A.~J., Belokurov V., Evans N.~W., 2011, MNRAS, 416, 2903 

\bibitem[\protect\citeauthoryear{Drake et al.}{2013a}]{drake13a} 
Drake A.~J., et al., 2013, ApJ, 763, 32 

\bibitem[\protect\citeauthoryear{Drake et al.}{2013b}]{drake13b} 
Drake A.~J., et al., 2013, ApJ, 765, 154

\bibitem[\protect\citeauthoryear{Eyer et al.}{2012}]{eyer12} 
Eyer L., et al., 2012, Ap\&SS, 341, 207

\bibitem[\protect\citeauthoryear{Freeman
\& Bland-Hawthorn}{2002}]{freeman02} Freeman K., Bland-Hawthorn J., 2002, ARA\&A, 40, 487

\bibitem[\protect\citeauthoryear{Frieman et
al.}{2008}]{frieman08} Frieman J.~A., et al., 2008, AJ, 135, 338

\bibitem[\protect\citeauthoryear{Ivezi{\'c} et 
al.}{2005}]{ivezic05} Ivezi{\'c} {\v Z}., Vivas A.~K., Lupton 
R.~H., Zinn R., 2005, AJ, 129, 1096

\bibitem[\protect\citeauthoryear{Ivezi{\'c} et 
al.}{2008}]{ivezic08} Ivezi{\'c} {\v Z}., et al., 2008, ApJ, 
684, 287 

\bibitem[\protect\citeauthoryear{Juri{\'c} et 
al.}{2008}]{juric08} Juri{\'c} M., et al., 2008, ApJ, 673, 864 

\bibitem[\protect\citeauthoryear{Kaiser et 
al.}{2002}]{kai2002} Kaiser N. et al., 2002, in Tyson J. A., Wolff S.,
  eds, Proc. SPIE Vol. 4836, Survey and Other Telescope Technologies
  and Discoveries. SPIE, Bellingham, p. 154

\bibitem[\protect\citeauthoryear{Keller et al.}{2008}]{keller08} 
Keller S.~C., Murphy S., Prior S., Da Costa G., Schmidt B., 2008, ApJ, 678, 
851 

\bibitem[\protect\citeauthoryear{Kinemuchi et 
al.}{2006}]{kinemuchi06} Kinemuchi K., Smith H.~A., Wo{\'z}niak 
P.~R., McKay T.~A., ROTSE Collaboration, 2006, AJ, 132, 1202

\bibitem[\protect\citeauthoryear{Klein et al.}{2014}]{klein14}
Klein C. R., Richards J. W., Butler N. R., Bloom J. S., 2014, MNRAS,
440, L96

\bibitem[\protect\citeauthoryear{Lafler 
\& Kinman}{1965}]{lafler65} Lafler J., Kinman T.~D., 1965, ApJS, 11, 216

\bibitem[\protect\citeauthoryear{Law 
\& Majewski}{2010}]{law10} Law D.~R., Majewski S.~R., 2010, ApJ, 714, 229

\bibitem[\protect\citeauthoryear{Liu et al.}{2013}]{liu13}
Liu X.-W., et al., 2013, in Feltzing S., Zhao G., Walton N., Whitelock
P., eds, Proc. IAU Symp. 298, Setting the scene for Gaia and LAMOST,
Cambridge University Press, pp. 310-321

\bibitem[\protect\citeauthoryear{Madore et al.}{2013}]{madore13} 
Madore B.F., et al., 2013, ApJ, 776, 135

\bibitem[\protect\citeauthoryear{Miceli et al.}{2008}]{miceli08} 
Miceli A., et al., 2008, ApJ, 678, 865 

\bibitem[\protect\citeauthoryear{Moody et 
al.}{2003}]{moody03} Moody R., Schmidt B., Alcock C., Goldader J., Axelrod T., Cook K.~H., Marshall S., 2003, EM\&P, 92, 125 

\bibitem[\protect\citeauthoryear{Newberg et 
al.}{2002}]{newberg02} Newberg H.~J., et al., 2002, ApJ, 569, 245 

\bibitem[\protect\citeauthoryear{Palaversa et 
al.}{2013}]{palaversa13} Palaversa L., et al., 2013, AJ, 146, 101

\bibitem[\protect\citeauthoryear{Rau et al.}{2009}]{rau2009}
Rau A., et al., 2009, PASP, 121, 1334

\bibitem[\protect\citeauthoryear{Saha 
\& Hoessel}{1990}]{saha90} Saha A., Hoessel J.~G., 1990, AJ, 99, 97

\bibitem[\protect\citeauthoryear{Sako et al.}{2008}]{sako08} 
Sako M., et al., 2008, AJ, 135, 348 

\bibitem[\protect\citeauthoryear{Schlegel, Finkbeiner, 
\& Davis}{1998}]{schlegel98} Schlegel D.~J., Finkbeiner D.~P., Davis M., 1998, ApJ, 500, 525

\bibitem[\protect\citeauthoryear{Schneider et 
al.}{2010}]{schneider10} Schneider D.~P., et al., 2010, AJ, 139, 
2360

\bibitem[\protect\citeauthoryear{Sesar et al.}{2007}]{sesar07} 
Sesar B., et al., 2007, AJ, 134, 2236

\bibitem[\protect\citeauthoryear{Sesar et al.}{2010}]{sesar10} 
Sesar B., et al., 2010, ApJ, 708, 717

\bibitem[\protect\citeauthoryear{Sesar et al.}{2011a}]{sesar11a} 
Sesar B., Stuart J.~S., Ivezi{\'c} {\v Z}., Morgan D.~P., Becker A.~C.,
Wo{\'z}niak P., 2011, AJ, 142, 190

\bibitem[\protect\citeauthoryear{Sesar et al.}{2011b}]{sesar11b}
Sesar, B., Juri{\'c}, M., \& Ivezi{\'c}, {\v Z}.\ 2011, \apj, 731, 4 

\bibitem[\protect\citeauthoryear{Sesar et al.}{2013}]{sesar13} 
Sesar B., et al., 2013, ApJ, 146, 21

\bibitem[\protect\citeauthoryear{Smith et al.}{2009}]{smith09a} 
Smith M.~C., et al., 2009, MNRAS, 399, 1223 

\bibitem[\protect\citeauthoryear{Soszy{\'n}ski et 
al.}{2009}]{soszyinsky09} Soszy{\'n}ski I., et al., 2009, AcA, 59, 1

\bibitem[\protect\citeauthoryear{Soszy{\~n}ski et 
al.}{2010}]{soszyinsky10} Soszy{\~n}ski I., Udalski A., 
Szyma{\~n}ski M.~K., Kubiak J., Pietrzy{\~n}ski G., Wyrzykowski {\L}., 
Ulaczyk K., Poleski R., 2010, AcA, 60, 165 

\bibitem[\protect\citeauthoryear{Soszy{\'n}ski et 
al.}{2011}]{soszyinsky11} Soszy{\'n}ski I., et al., 2011, AcA, 61, 1 

\bibitem[\protect\citeauthoryear{Udalski et al.}{1992}]
{udalski92} Udalski A., Szymanski M., Kaluzny J., 
Kubiak M., Mateo M., 1992, AcA, 42, 253

\bibitem[\protect\citeauthoryear{Vivas et al.}{2004}]{vivas04} 
Vivas A.~K., et al., 2004, AJ, 127, 1158 

\bibitem[\protect\citeauthoryear{Vivas
\& Zinn}{2006}]{vivas06} Vivas A.~K., Zinn R., 2006, AJ, 132, 714 

\bibitem[\protect\citeauthoryear{Watkins et 
al.}{2009}]{watkins09} Watkins L.~L., et al., 2009, MNRAS, 398, 
1757 

\bibitem[\protect\citeauthoryear{Xue et al.}{2008}]{xue08} 
Xue X.~X., et al., 2008, ApJ, 684, 1143 

\bibitem[\protect\citeauthoryear{Yanny et al.}{2003}]{yanny03} 
Yanny B., et al., 2003, ApJ, 588, 824

\bibitem[\protect\citeauthoryear{York et al.}{2000}]{york00} 
York D.~G., et al., 2000, AJ, 120, 1579

\bibitem[\protect\citeauthoryear{Yuan, Liu, \& Xiang}{2013}]{yuan13}
Yuan H.-B., Liu X.-W., Xiang M.-S., 2013, arXiv, arXiv:1301.1427 

\bibitem[\protect\citeauthoryear{Zhang et al.}{2013}]{zhang13}
Zhang H.-H., et al., 2013, RAA, 13, 490

\bibitem[\protect\citeauthoryear{Zhang et al.}{2014}]{zhang14}
Zhang H.-H., et al., 2014, RAA, 14, 456

\end{thebibliography}
\end{document}